\documentclass{aa}  

\usepackage{amsmath}
\usepackage{graphicx}
\usepackage{graphics}
\usepackage{txfonts}
\usepackage{natbib}
\usepackage{xcolor}
\usepackage{url}

\bibpunct{(}{)}{;}{a}{}{,}

\def\teff{\ifmmode T_{\rm eff} \else $T_{\mathrm{eff}}$\fi}

\def\ltsima{$\buildrel<\over\sim$}
\def\lsim{\lower.5ex\hbox{\ltsima}}

\newcommand{\hii}{H~{\sc ii}}
\newcommand{\ha}{\ifmmode {\rm H}\alpha \else H$\alpha$\fi}
\newcommand{\hb}{\ifmmode {\rm H}\beta \else H$\beta$\fi}
\newcommand{\hg}{\ifmmode {\rm H}\gamma \else H$\gamma$\fi}
\newcommand{\lya}{\ifmmode {\rm Ly}\alpha \else Ly$\alpha$\fi}

\newcommand{\heii}{He~{\sc ii}}
\newcommand{\Heiiuv}{He~{\sc ii} $\lambda$1640}

\newcommand{\ebv}{\ifmmode E_{\rm B-V} \else $E_{\rm B-V}$\fi}
\newcommand{\av}{\ifmmode A_{\rm V} \else $A_{\rm V}$\fi}

\def\cmc{cm$^{-3}$}

\def\ergs{erg s$^{-1}$}

\def\msun{\ifmmode M_{\odot} \else M$_{\odot}$\fi}
\def\msunyr{\ifmmode M_{\odot} {\rm yr}^{-1} \else M$_{\odot}$ yr$^{-1}$\fi}
\def\zsun{\ifmmode Z_{\odot} \else Z$_{\odot}$\fi}

\def\lsun{\ifmmode L_{\odot} \else L$_{\odot}$\fi}

\def\mup{\ifmmode M_{\rm up} \else M$_{\rm up}$\fi}
\def\mlow{\ifmmode M_{\rm low} \else M$_{\rm low}$\fi}

\def\aap{A\&A}

\def\apjl{ApJL}
\def\apjs{ApJS}

\newcommand{\oh}{\ifmmode 12 + \log({\rm O/H}) \else$12 + \log({\rm
O/H})$\fi}

\newcommand{\civ}{C~{\sc iv}}

\def\Nii{[N\small II]\normalsize $\lambda\lambda$6548,6584}
\def\Sii{[S~{\sc ii}] $\lambda\lambda$6717,6731}

\def\Oii{[O~{\sc ii}] $\lambda$3727}
\def\Oiii{[O~{\sc iii}] $\lambda\lambda$4959,5007}

\def\flyf{\ifmmode f_{\rm Lyf} \else $f_{\rm Lyf}$\fi}
\def\pz{\ifmmode P(z) \else $P(z)$\fi}
\def\ki2{\ifmmode \chi^2 \else $\chi^2$\fi}
\def\zphot{\ifmmode z_{\rm phot} \else $z_{\rm phot}$\fi}

\newcommand{\xphot}{\ifmmode x_\gamma \else $v_\gamma$\fi}
\newcommand{\xobs}{\ifmmode x_{\rm obs} \else $x_{\rm obs}$\fi}
\newcommand{\xcmf}{\ifmmode x_{\rm CMF} \else $x_{\rm CMF}$\fi}
\newcommand{\vexp}{\ifmmode V_{\rm exp} \else $V_{\rm exp}$\fi}
\newcommand{\vmax}{\ifmmode V_{\rm max} \else $V_{\rm max}$\fi}
\newcommand{\nh}{\ifmmode N_{\rm HI} \else $N_{\rm HI}$\fi}
\newcommand{\dv}{\ifmmode \Delta v({\rm em-abs}) \else $\Delta v({\rm em}-{\rm abs})$\fi}

\def\fesc{\ifmmode f_{\rm esc} \else $f_{\rm esc}$\fi}
\def\fescrel{\ifmmode f_{\rm esc,rel} \else $f_{\rm esc,rel}$\fi}

\def\frellya{\ifmmode f^{\rm rel}_{\rm{Ly}\alpha} \else $f^{\rm rel}_{\rm{Ly}\alpha}$\fi}

\def\hii{H{\sc ii}}

\newcommand{\mstar}{\ifmmode M_\star \else $M_\star$\fi}
\newcommand{\muv}{\ifmmode M_{1500} \else $M_{1500}$\fi}
\newcommand{\auv}{\ifmmode A_{\rm UV} \else $A_{\rm UV}$\fi}
\newcommand{\luv}{\ifmmode L_{\rm UV} \else $L_{\rm UV}$\fi}
\newcommand{\lir}{\ifmmode L_{\rm IR} \else $L_{\rm IR}$\fi}
\newcommand{\lbol}{\ifmmode L_{\rm bol} \else $L_{\rm bol}$\fi}
\newcommand{\liruv}{\ifmmode L_{\rm IR+UV} \else $L_{\rm IR+UV}$\fi}
\newcommand{\liroveruv}{\ifmmode L_{\rm IR}/L_{\rm UV} \else $L_{\rm IR}/L_{\rm UV}$\fi}
\newcommand{\nlyc}{\ifmmode N_{\rm Lyc} \else $N_{\rm Lyc} $\fi}
\newcommand{\rholyc}{\ifmmode \rho_{\rm Lyc} \else $\rho_{\rm Lyc} $\fi}
\newcommand{\chion}{\ifmmode \xi_{\rm ion} \else $\xi_{\rm ion}$\fi}
\newcommand{\chioncorr}{\ifmmode \xi_{\rm ion}^0 \else $\xi_{\rm ion}^0$\fi}
\newcommand{\kappauv}{\ifmmode \kappa_{\rm UV} \else $\kappa_{\rm UV}$\fi}
\newcommand{\kappaunits}{\ifmmode {\rm M}_\odot {\rm yr}^{-1} / ({\rm erg \,s}^{-1} {\rm Hz}^{-1}) \else ${\rm M}_\odot {\rm yr}^{-1} / ({\rm erg\, s}^{-1} {\rm Hz}^{-1})$\fi}

\newcommand{\qh}{\ifmmode Q_H \else $Q_H$\fi}

\newcommand{\Civuv}{C~{\sc iv} $\lambda$1550}

\newcommand{\Ciii}{C~{\sc iii}]}
\newcommand{\Ciiiuv}{C~{\sc iii}] $\lambda$1909}

\newcommand{\bpass}{{\em BPASS}}

\begin{document}

\title{Observable and ionizing properties of star-forming galaxies with very massive stars and different initial mass functions}
\subtitle{}
\author{D. Schaerer\inst{1,2}, 
J. Guibert\inst{1}, 
R. Marques-Chaves\inst{1}, 
F. Martins\inst{3}
}
  \institute{Observatoire de Gen\`eve, Universit\'e de Gen\`eve, Chemin Pegasi 51, 1290 Versoix, Switzerland
         \and
CNRS, IRAP, 14 Avenue E. Belin, 31400 Toulouse, France
        \and
LUPM, Universit\'e de Montpellier, CNRS, Place Eug\`ene Bataillon, F-34095 Montpellier, France
         }

\authorrunning{D.\ Schaerer et al.}
\titlerunning{}

\date{Accepted for publication in A\&A}

\abstract{The presence of very massive stars (VMS) with masses $>100$ \msun\ is now firmly established in the Local Group, nearby galaxies, and out to cosmological distances. If present, these stars could boost the UV luminosity and ionizing photon production of galaxies, helping  to alleviate the overabundance of UV-bright galaxies found with  JWST at high redshift.}
{To examine these questions,
we quantify the impact of VMS on properties of integrated stellar populations, exploring different stellar initial mass functions (IMFs) extending up to 400 \msun, and with slopes between standard (Salpeter-like) and flatter, more top-heavy IMFs.}
{Combing consistent stellar evolution and atmosphere models tailored to VMS 
at $1/2.5$ \zsun\ metallicity with \bpass\ evolutionary synthesis models and including nebular emission, we computed integrated spectral energy distributions (SEDs) and derived quantities for a large set of models.}
{We find that VMS contribute significantly to the UV luminosity and Lyman continuum emission of young stellar populations, and they are characterized by strong stellar  \heii\ emission, with EW(\Heiiuv) up to 4--8 \AA\ at young ages or $\sim 2.5-4$ \AA\ for a constant star formation rate (SFR) (for the IMFs considered here).
For IMFs with a Salpeter slope, the boost of the UV luminosity is relatively modest (up to a factor of $\sim 1.6$). However, small changes in the IMF slope (e.g., from $\alpha_2=-2.35$ to $-2$)  lead to large increases in \luv\ and the ionizing photon production, $Q$.
The ionizing photon efficiency, \chion, is also increased with VMS, by typically 0.14-0.2 dex for a Salpeter slope, and by up to $\sim 0.4$ dex when the IMF is extended up to 400 \msun. Stronger H recombination lines are also predicted in the presence of VMS.
Interestingly, SEDs including VMS show smaller Lyman breaks, and the shape of the ionizing spectra remain essentially unaltered up to $\sim 35$ eV, but become softer at higher energies.
We derive and discuss the maximum values that quantities such as \luv\ per stellar mass or unit SFR, \chion, $Q$, and others can reach when VMS are included, and we show that these values become essentially independent of the IMF.
We propose  observational methods to test for the presence of VMS and constrain the IMF in star-forming galaxies. 
Finally, using published JWST observations, we examine if high redshift ($z \protect\ga 5-6$) galaxies show some evidence of the presence of VMS and/or signs of non-standard IMFs. Very top-heavy IMFs can be excluded on average, but we find that the IMF could well extend into the regime of VMS and be flatter than Salpeter in the bulk of high-$z$ galaxies.
}
{The  predictions should improve our understanding of the stellar content and IMF of the most distant galaxies and allow us to establish if ``extreme'' or ``unusual'' populations extending well beyond 100 \msun\ existed in the early Universe.}

 \keywords{Galaxies: stellar content -- Galaxies: high redshift -- Galaxies: ISM -- Cosmology: dark ages, reionization, first stars}

 \maketitle

\section{Introduction}
\label{s_intro}

Stars with masses of $\mstar \sim 10-100$ \msun\ (generally referred to as massive stars) are ubiquitous in the nearby and distant Universe. They power \hii\ regions and star-forming galaxies and their respective radiative output. Together with supernovae, massive stars dominate the mechanical feedback and chemical evolution in star-forming galaxies during the first gigayear of their evolution \citep{Conti2008From-Luminous-H}. In recent years, the existence of stars with initial masses higher than 100 \msun\ -- so-called very massive stars (VMS, \citealt{Vink2015Very-Massive-St}) -- has been established or suggested, both in the Local Group and at cosmological distances, but their general impact on star-forming galaxies remains largely to be explored. This is one of main aims of the present work.

The best evidence for individual VMS comes from the well-studied region R136 at the core of the giant \hii\ region 30 Dor in the Large Magellanic Cloud, which is resolved into hundreds of massive OB stars and several peculiar stars (with spectral types WNh), whose masses have now been established between 100-200 \msun\ \citep[e.g.,][]{Crowther2010The-R136-star-c,Bestenlehner2020The-R136-star-c,Kalari2022Resolving-the-C}.
In R136, a very young region (age $\sim 1\pm 1$ Myr), the seven VMS produce a strong \Heiiuv\ emission line (with EW$\sim$4~\AA), and they are found to contribute 32\% of the far-UV continuum (18\% of the entire Tarantula nebula) and $\sim 25$ \% of the ionizing photon production \citep{Crowther2016The-R136-star-c,Bestenlehner2020The-R136-star-c,Crowther2024Mapping-the-cor}.

The presence of VMS is also fairly well established or suspected in several nearby star-forming regions, young clusters, or galaxies, as is testified by their strong \Heiiuv\ emission and other indirect evidence \citep{Leitherer2018Physical-Proper,Smith2016The-Very-Massiv,Senchyna2021Ultraviolet-spe,Wofford2023Extreme-broad-H,Smith2023HST-FUV-Spectro}.
 A recent search and critical analysis of VMS in nearby objects and low-$z$ galaxies has been presented by \cite{Martins2023Inferring-the-p}. 
 
 At cosmological distances, signatures of VMS have possibly been found in a gravitationally lensed star cluster in the Sunburst arc at $z=2.37$ \citep[][]{Mestric2023Clues-on-the-pr,Rivera-Thorsen2024The-Sunburst-Ar}, and \cite{Upadhyaya2024Evidence-for-ve} have found evidence for VMS in very UV-bright (\muv $\sim -23$ to $-24.5$) compact star-forming galaxies discovered in the Sloan Digital Sky Survey by \cite{Marques-Chaves2020The-discovery-o,Marques-Chaves2021The-UV-brightes,Marques-Chaves2022An-extreme-blue}.
%
Interestingly, \cite{Upadhyaya2024Evidence-for-ve} find that galaxies showing signatures of VMS are frequent (found in $\sim 69$ \%) of UV-bright galaxies, whereas the presence of VMS is quite rarely seen otherwise (e.g., in $\la 1$\% of LBGs at $z \sim 2-5$).  This suggests that the presence of VMS, and an hence an extension of the IMF beyond the classical upper mass limit of $\mup \sim 100-120$ \msun,  might be more common or even frequent in high redshift star-forming galaxies.

In parallel, early observations with the JWST have revealed an unexpectedly large number of UV-luminous galaxies with blue spectral slopes in the early Universe  \citep[$z \ga 5-7$, see e.g.,][]{Topping2024The-UV-Continuu,Cullen2024The-ultraviolet}.
To explain this tension, several physical scenarios have been proposed, including dust clearing by radiation-driven outflows \citep{Ferrara2023On-the-stunning}, a higher star formation efficiency, possibly due to reduced feedback \citep{Dekel2023Efficient-forma},  stochastic star formation histories that can lead to biases boosting the UV luminosity \citep{Mason2023The-brightest-g,Shen2023The-impact-of-U}, and different IMFs favoring massive stars \citep{Finkelstein2023CEERS-Key-Paper,Harikane2023A-Comprehensive,Trinca2024Exploring-the-n,Woodrum2023JADES:-Using-NI}.
For all these reasons, examining the role and impact of VMS and the effect of different IMFs on the observable properties of star-forming galaxies is of interest in a variety of contexts. This is the main goal of the this paper.

To model the spectral evolution of stellar populations including VMS, it is important to consistently compute stellar evolution and atmosphere models with appropriate prescriptions for stellar mass loss and wind properties, and, ideally, to test and/or ``calibrate'' the models against observations. Several synthesis models already allow for extensions of the IMF beyond 100 \msun, into the regime of VMS. For example, both the BPASS and the updated Charlot \& Bruzual models  contain predictions for IMFs up to 100 and 300 \msun\ \citep[see][and references therein]{Eldridge2017Binary-Populati,BC03,Plat2019Constraints-on-,Mayya2023Detection-of-He}. However, both model sets show difficulties in reproducing several UV observations and fail to reproduce in particular the most important spectral diagnostic of VMS, the \Heiiuv\ emission, as has been shown by \cite{Senchyna2021Ultraviolet-spe}, \cite{Martins2023Inferring-the-p}, and \cite{Smith2023HST-FUV-Spectro}. The main reason for this failure is most likely the lack of appropriate atmosphere models for VMS, since the models cited before treat these objects with classical WM-BASIC models developed for O stars. 
Combining stellar evolution and non-LTE atmosphere models including strong stellar winds, \cite{Martins2022Spectroscopic-e} have been able to reproduce the spectra of individual VMS in the LMC, the integrated spectrum of R136, and they also reproduce UV spectra of other clusters with strong \heii\ features. These VMS models, combined with BPASS for stars $\le 100$ \msun, have also been successfully compared to the observations of the Sunburst arc cluster at $z=2.37$ and to a sample of UV-bright galaxies at $z\sim 2-4$, which show UV spectra indicative of the presence of VMS \citep{Mestric2023Clues-on-the-pr,Upadhyaya2024Evidence-for-ve}.

In this work, we use these VMS models to explore in more detail the UV, ionizing, and related observable properties of stellar populations including stars in the range of $\sim 100-400$ \msun. We compute in particular the UV and Lyman continuum emission of such populations, including the associated nebular emission (primarily nebular continuum and H recombination lines), for a range of IMFs, covering different upper and lower mass limits and two different IMF slopes (Salpeter and a flatter IMF).  We then show how the UV luminosity, the ionizing photon production and efficiency, line equivalent widths, and other properties are ``boosted'' by the presence of VMS and how they depend on the IMF. We also examine the shape of the ionizing spectra of populations including VMS, showing that they have smaller Lyman breaks and softer spectra than normal populations.
To understand what maximum predictions are possible, e.g.~for the UV luminosity per stellar mass or star formation rate, for the ionizing photon efficienciency, \chion, and others, we also consider integrated populations completely devoid of normal stars (i.e., populations of pure VMS), and we show that most observational properties are naturally bounded by some maximum values, which we derive. We also briefly discuss cases dominated by pure nebular continuum emission. 
The model predictions presented here are made available publicly and should hopefully contribute to a better understanding of the importance of VMS and effects of unusual IMFs in star-forming galaxies.

In Sect.\ \ref{s_models}, we briefly describe the inputs of our model calculations.
The predictions of our model grids are shown and discussed in Sect.~\ref{s_pred}.
In Sect.~\ref{s_discuss}, we compare our work with other studies, discuss the maximum UV and ionizing photon efficiencies of populations including VMS,
present observational tests for extreme IMFs, and compare our results to recent observation from  JWST.
Our main results are summarized in Sect.\ \ref{s_conclude}.

\begin{table*}[htp]
\caption{Overview of the model grids and their main parameters, computed with and without nebular  continuum emission, for metallicity $Z=0.006$}
\begin{center}
\begin{tabular}{llllll}
Model ID &  Models used & IMF slope $\alpha_2$ & \mup& \mlow & $M_{\rm VMS}/M_{\rm tot}$\\
 & & &  [\msun] &  [\msun] \\
\hline \\
BPASS-100 & BPASS           &  $-2.35, -2.0$ & 100 & 0.1 & 0. \\
BPASS-300 & BPASS           &  $-2.35, -2.0$ & 300 & 0.1 & 0.043 \\
VMS-150     & BPASS+VMS  &  $-2.35, -2.0$ & 150 & 0.1 & 0.015, 0.055\\
VMS-200     & BPASS+VMS  &  $-2.35, -2.0$ & 200 & 0.1 & 0.028, 0.11 \\
VMS-300     & BPASS+VMS  &  $-2.35, -2.0$ & 300 & 0.1 & 0.043,  0.17 \\
VMS-400     & BPASS+VMS  &  $-2.35, -2.0$ & 400 & 0.1 & 0.051, 0.22 \\
VMS-only-150    & VMS               &  $-2.35$         & 150 & 100 & 1. \\
VMS-only-400    & VMS               &  $-2.35$         & 400 & 100  & 1.\\

\end{tabular}
\end{center}
\label{ta_models}
\end{table*}

\section{Models}
\label{s_models}

To study the impact of VMS on the integrated properties of star-forming regions and galaxies we use the recent
combined stellar evolution and atmosphere models of \cite{Martins2022Spectroscopic-e} at 1/2.5 solar metallicity.
The models include the specific mass loss rates of VMS, that is mass loss rates that are boosted compared to normal O-type stars \citep{graef08,vink11,besten20} and that are the main ingredient leading to the special spectroscopic appearance of VMS. The mass loss recipe for VMS was adopted from \citep{graef21}. It relies on a calibration anchored on the observed wind properties of the VMS in R136 at the core of the Tarantula nebula in the LMC. Stellar tracks were computed for masses between 150 and 400~\msun\ with the code \emph{STAREVOL} \citep{amard}. Atmosphere models and synthetic spectra were subsequently calculated along each track, at positions corresponding to ages between 0 and 2.5~Myr (with steps of 0.5~Myr). The atmosphere models were computed with the code \emph{CMFGEN} \citep{hm98} that include the physics relevant for massive stars (winds, spherical geometry, line-blanketing, non-local thermodynamical equilibrium). The evolution and atmosphere models are consistent in terms of stellar properties (effective temperature, mass, radius, luminosity, surface chemical composition): the output of the evolutionary calculations are taken for input in the atmosphere calculations. The final synthetic spectra sample the main sequence of VMS. 
The optical spectra resemble those of O supergiants with strong winds (OIf stars), transition objects (OIf/WN), or hydrogen-rich WN (WNh) stars. In the UV, the spectra show the classical P-Cygni emission lines (including \Civuv) and several features unique to VMS, in particular strong broad \Heiiuv\ emission \citep{Martins2022Spectroscopic-e}.
The late stages of stellar evolution, corresponding to about 10\% of the star's lifetime, were not considered for the reasons explained in \cite{Martins2022Spectroscopic-e}. 

The VMS models are added to ``normal" stars, defined here as stars with initial masses $\le 100$ \msun, making use of the recent
\bpass\ evolutionary synthesis models of \cite{Eldridge2017Binary-Populati}, following the simple procedure adopted by 
\cite{Martins2022Spectroscopic-e} and updates described in \cite{Upadhyaya2024Evidence-for-ve}.
 In practice, we use the \bpass\ models v.2.2.1, including binary stars for a metallicity $Z=0.006$, 
the closest available to $Z=0.00536=1/2.5$ \zsun\ for $\zsun = 0.0134$ adopted  by \cite{Martins2022Spectroscopic-e}. 
We explore different stellar initial mass functions, including the ``classical'' IMF, described by a Salpeter slope of $\alpha_2=-2.35$ for masses $\mstar \ge 0.5$ \msun\ and $\alpha_1=-1.3$ at low masses (from the lower mass limit of $\mlow=0.1$ to 0.5 \msun) and a flatter IMF ($\alpha_2=-2.0$) giving more weight to massive stars. The maximum stellar mass considered varies between $\mup=100$ and 400 \msun, and we also consider extreme cases with populations formed exclusively with very massive stars; that is, $\mlow=100$ \msun\ (VMS-only in the following).
The different models used in this study and their parameters are summarized in Table \ref{ta_models}.

Since nebular emission is significant for populations including young and massive stars we include nebular continuum emission (free-bound, free-free, and two-photon) from H and He and added to the stellar spectra. We discuss our assumptions  for the nebular continuum below.
Finally, spectral energy distributions (SEDs) are computed for two limiting star formation histories, namely instantaneous bursts
and constant star formation rates (SFR). Our models are normalized to a total stellar mass of $10^6$ \msun\ for bursts or to ${\rm SFR}=1$ \msunyr.
A selection of the main quantities predicted by our models are tabulated in Table \ref{ta_predict} below. 
Detailed model predictions, including SEDs, are available on request from the first author and made available electronically.

To compute nebular continuum emission (free-bound, free-free, and two-photon) from H and He we use {\tt pyneb} \citep{Morisset2020Atomic-Data-Ass}, assuming assume  electron temperatures and densities ($T_e=15000$ K, $n_e=100$ \cmc) typical of high redshift moderately low-metallicity sources  \citep[e.g.,][]{Sanders2020The-MOSDEF-surv,Curti2023The-chemical-en}. 
For significantly lower metallicities ($\la 0.03-0.1$ solar) than adopted here, the nebular continuum can in principle be enhanced compared to case B assumption by a factor proportional to the mean energy of the ionizing photons, due to enhanced photoionization rates as shown by \cite{raiter2010} and discussed recently by \cite{Katz202421-Balmer-Jump-}. 
Finally, the effect of varying electron temperatures are illustrated in the Appendix (see Fig.~\ref{fig_nebcont}). Significant increases in the  electron density beyond $n_e \ga 10^{3-4}$ \cmc\ decrease the two-photon emission, in other words nebular emission in the rest-UV, but have little effect at longer wavelengths. For a more detailed discussion of these effects, which may thus affect individual objects especially at very low metallicity we refer, for example, to \cite{Bottorff2006Two-Photon-Tran,raiter2010,Katz202421-Balmer-Jump-}. However, since only  few galaxies observed with JWST show metallicities  $\la 0.03-0.1$ solar, we do not expect significantly enhanced two-photon emission in the majority of them, and hence consider our assumptions should indeed be quite representative of  ``typical'' objects with the moderately low-metallicity, adopted here.

One of the limitations of this work originates from the availability of detailed stellar evolution and atmosphere models for VMS, which are so far essentially only available for a single metallicity. In this sense our work should be considered  as  a pilot study, before a broader metallicity range can be explored. 
We acknowledge that our models are tailored for the LMC metallicity. VMS are mostly characterized by their strong stellar winds that are only calibrated at that metallicity. There is so far no direct information about the behavior of VMS winds or calibration of VMS mass loss rates at lower metallicity. Some empirical studies argue that VMS mass loss rates do not change with $Z$ \citep{Smith2023HST-FUV-Spectro}, while some theoretical prediction of H-free WR stars call for complex metallicity dependence \citep{Sander2020On-the-nature-o}. 
The spectroscopic evolution of VMS at low $Z$ will be the topic of a subsequent publication (Martins et al., in prep). We briefly discuss this issue further in Sect.~\ref{s_caveats}.

\begin{figure*}[htb]
\centering\
 \includegraphics[width=0.45\textwidth]{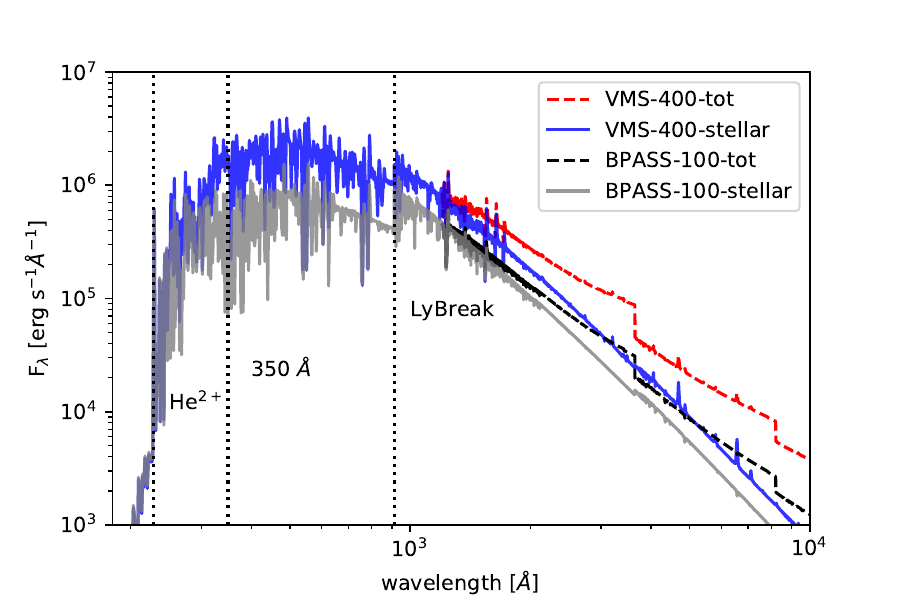}
 \includegraphics[width=0.45\textwidth]{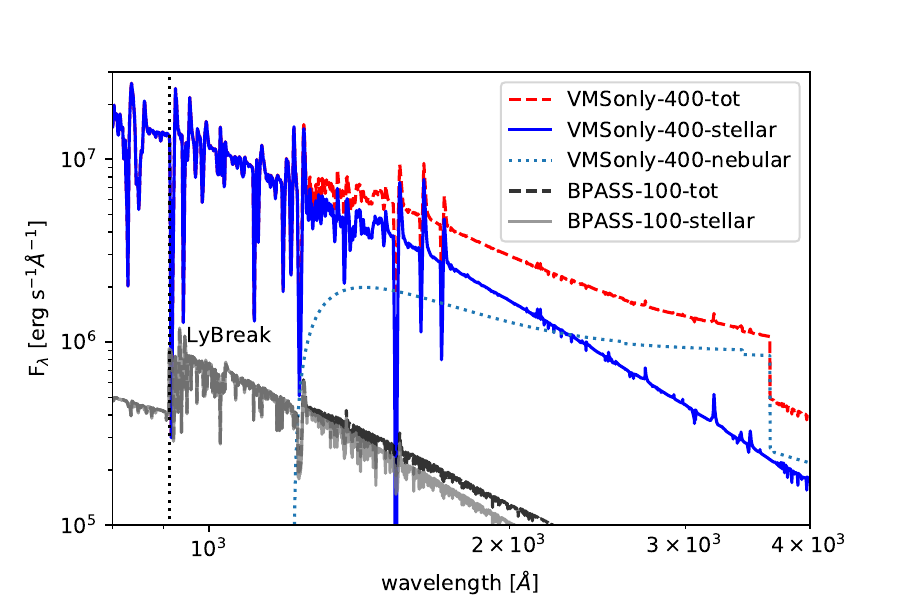}
\caption{Comparison of predicted SEDs for a 1 Myr old population with and without VMS.
{\bf Left panel:} 
The standard BPASS model with $\mup=100$ \msun\ is shown in gray (stellar emission only) and in black (including nebular continuum emission).
Blue and red lines show the model extending up to 400 \msun, the former including only stellar emission, the latter the total continuum (stellar plus nebular continuum). 
{\bf Right panel:}  zoom on the UV domain, comparison the standard BPASS model with a population including only VMS (from 100 to 400 \msun).
The dotted line shows the nebular continuum emission of the VMS-only-400 population.}
\label{fig_seds}
\end{figure*}


\section{Predicted UV and ionizing properties of populations including VMS}
\label{s_pred}

To illustrate the impact of VMS on the UV--optical SED of integrated stellar populations, we show some examples from our models at an age of 1 Myr in Fig.~\ref{fig_seds}. In the left panel, which compares the SED of a populations with a Salpeter IMF extending up to $\mup=100$ and 400 \msun, one can clearly see the main changes due to VMS:
First, VMS boost the UV luminosity of stellar populations, both in the Lyman continuum (LyC, at $\lambda < 912$ \AA) and in the non-ionizing UV ($\lambda > 912$ \AA).
Second, the  Lyman break of population including VMS is smaller, in other words the ratio of LyC to non-ionizing flux is increased. 
Third, the ionizing spectrum of populations with VMS decreases more rapidly between 350 and 228 \AA, which means that their ionizing spectra are softer between $\sim 35$ and 54 eV.
These properties will be discussed further and quantified in what follows.

The right panel of Fig.~\ref{fig_seds} shows the SED of the VMS-only model, populated exclusively with masses between 100 and 400 \msun, compared to the  model with a classical IMF with $\mup = 100$ \msun, where in both cases the total stellar mass is identical ($10^6$ \msun, by construction).  In this case the UV luminosity boost is very high -- more than a factor 10 at most wavelengths -- and the Lyman break of the VMS-only-400 model is essentially absent. In the right panel we also show separately the nebular continuum, which obviously provides a significant contribution longward of \lya\ (at $\lambda > 1216$ \AA), as well known for populations with strong Lyman continuum emission \citep[see e.g.,][]{schaerer2002,schaerer2003,schaerer&debarros2011}.

\begin{figure}[tb]
\centering
\includegraphics[width=0.45\textwidth]{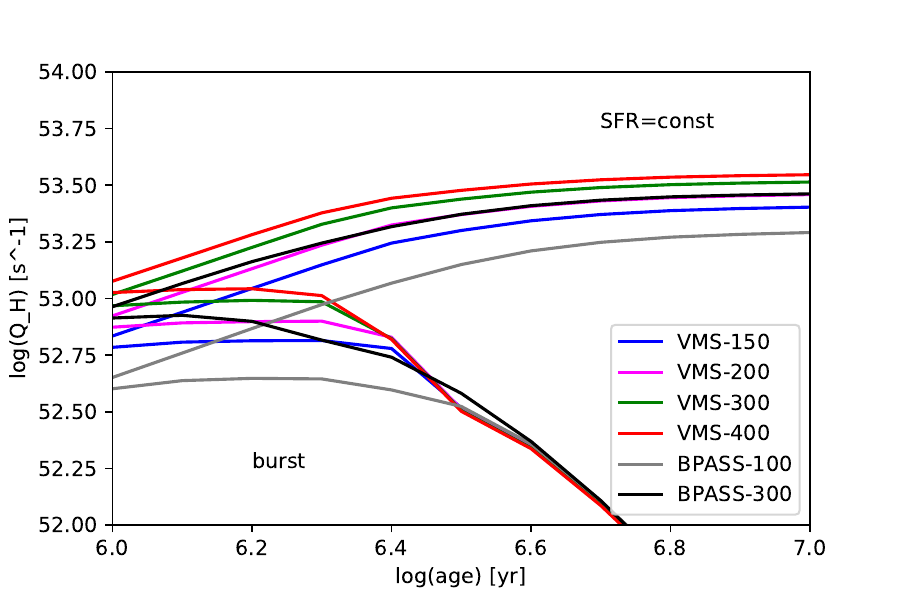}
\caption{Predicted temporal evolution of the ionizing photon production, \qh, of different models, for instantaneous bursts and constant SFR. 
Burst models are normalized to $10^6$ \msun, models for constant SFR  to  SFR$=1$ \msunyr.}
\label{fig_qh}
\end{figure}

\begin{figure}[tb]
\centering
\includegraphics[width=0.45\textwidth]{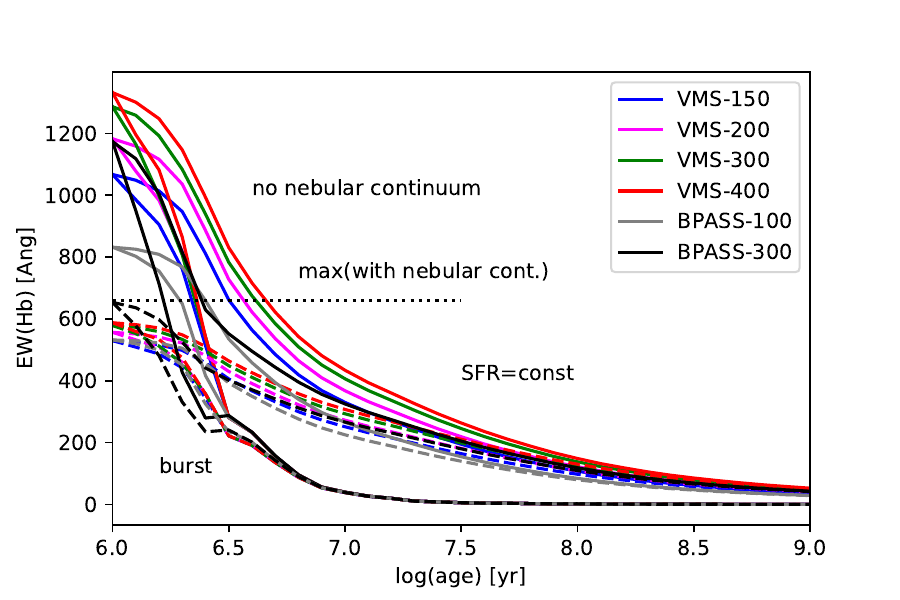}
\caption{Predicted temporal evolution of the \hb\ equivalent width of different models, for instantaneous bursts and constant SFR. 
Solid (dashed) lines show models without (with) nebular continuum emission.}
\label{fig_ew}
\end{figure}

\subsection{Ionizing photon flux and recombination line strengths}
The Lyman continuum photon production, \qh, which measures the number of H-ionizing photons emitted per unit time,  of the integrated populations including VMS is shown in Fig.~\ref{fig_qh} as a function of time and for the different maximum masses \mup\ considered.
Overall we see that an extension of the IMF beyond 100 \msun\ leads, as expected, to a stronger LyC flux per unit stellar mass or SFR. 
The increase in \qh\ is up to $\sim 0.4$ dex (a factor $\sim 2.5$) at young ages when stars up to 400 \msun\ are included. At equilibrium, in other words at ages $\ga 5$ Myr, the increase is of $\sim 0.2$ dex (factor $\sim 1.6$).  
We note also that the increase predicted by the calculations including the VMS models of \cite{Martins2022Spectroscopic-e} is somewhat larger than that predicted by the \bpass\ models which extend to 300 \msun. This  is due to differences in the stellar evolution and atmosphere models used by these works.

One of the immediate implications of the increased ionizing photon emission \qh\ is that SFR indicators using H recombination lines (e.g., \ha) must be revised if VMS are present. Since SFR$(\ha) = c_{\ha} \times L(\ha)$ and $L(\ha) \propto \qh$ (see also Sect.~\ref{s_appendix}), the conversion factor $c_{\ha}$ is simply increased by the amounts discussed above.

The predicted strength of the \hb\ recombination line, as measured by its equivalent width, EW(\hb), is shown in Fig.~\ref{fig_ew}.
It shows a well-known monotonic decrease with age due to the combined decrease in the ionizing photon flux and the increasing  flux in the optical domain. A maximum EW(\hb)$\approx 700$ \AA\ is reached by models including VMS up to 400 \msun, a factor 1.3 higher than for \mup=100 \msun\ in the models accounting for nebular  continuum emission. This ``boost'' is less than the increase in \qh, since $EW \propto \qh / L_{4861}$ and since both \qh\ and the continuum luminosity $L_{4861}$ at \hb\ wavelength increase in the presence of VMS.

Indeed, as well known the nebular continuum contributes a significant fraction of the flux in the optical domain for young populations (see Fig.~\ref{fig_seds}). If this contribution is artificially left out, the \hb\ equivalent width increases by $\sim 60-70$\%, as shown by the models with no nebular continuum in  Fig.~\ref{fig_ew}.
  
\begin{figure}[tb]
\centering
 \includegraphics[width=0.45\textwidth]{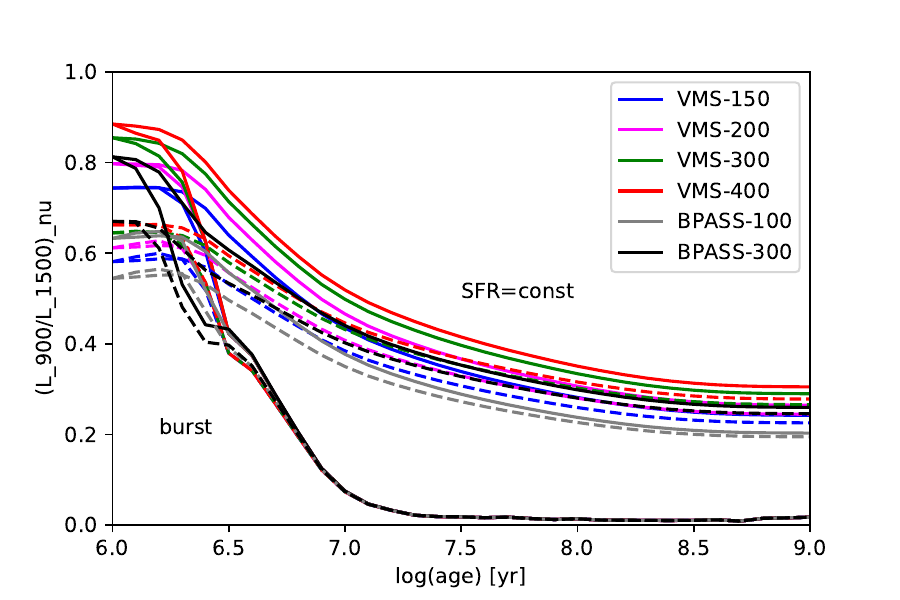}
\caption{Predicted temporal evolution of the $L_{900}/L_{1500}$ ratio of different models, for instantaneous bursts and constant SFR. 
Solid (dashed) lines show models without (with) nebular continuum emission.}
\label{fig_lyb}
\end{figure}

\begin{figure}[tb]
\centering
 \includegraphics[width=0.45\textwidth]{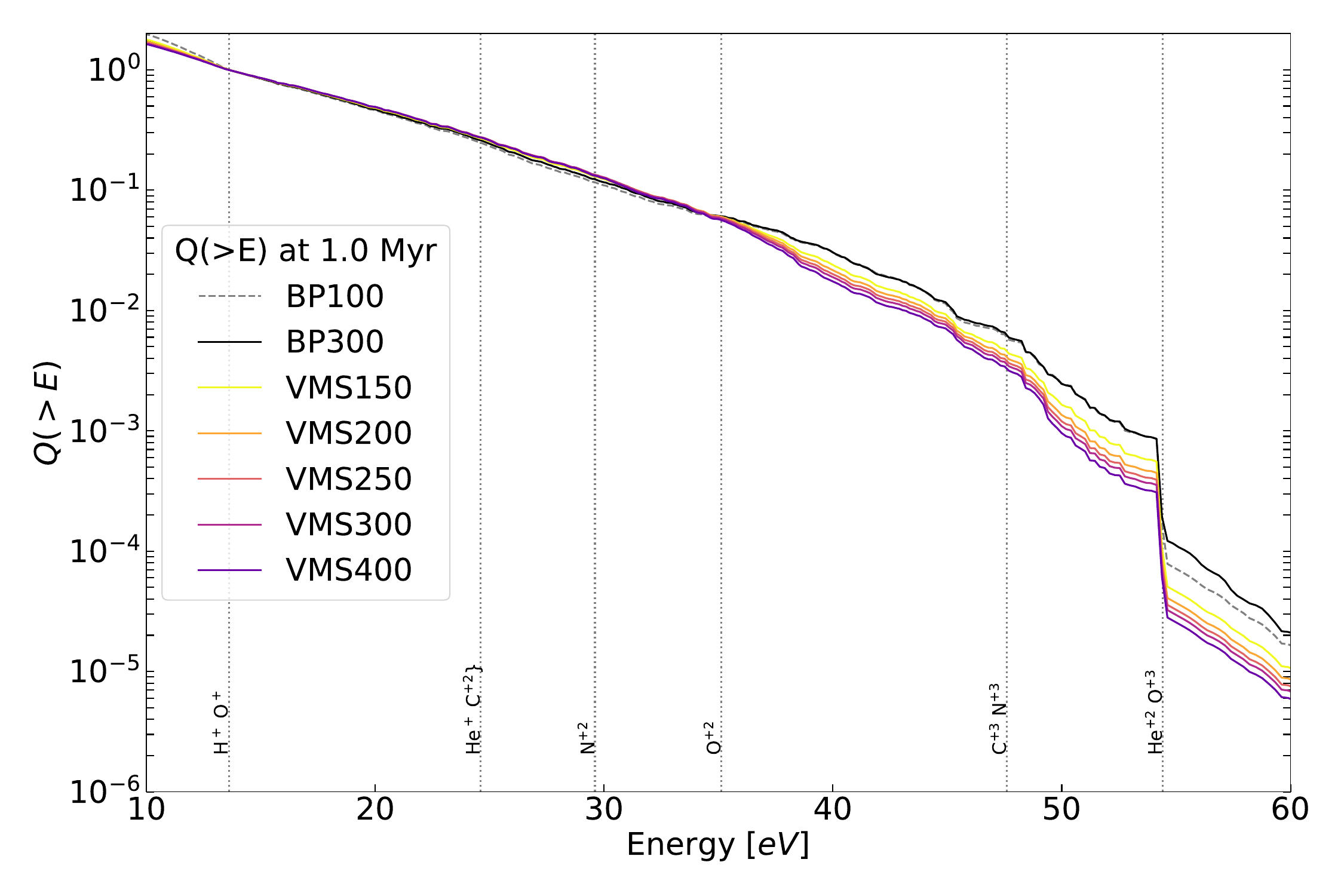}
\caption{Comparison of the ionizing spectra of different models for age 1 Myr. Shown is the cumulative number of ionizing photons emitted above energy $E$, $Q(>E)$,
as a function of the energy (in eV). All ionizing spectra are normalized to unity at the Lyman limit (13.6 eV), to allow for easy comparison of their relative hardness. Vertical lines mark the ionization potentials of several important ions.}
\label{fig_qe}
\end{figure}

\begin{figure*}[tb]
\centering
 \includegraphics[width=0.45\textwidth]{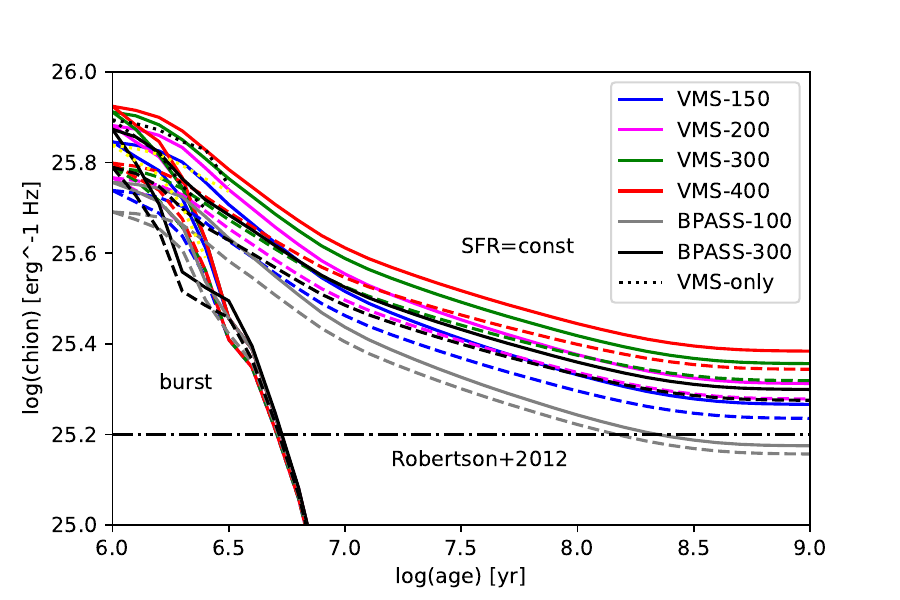}
 \includegraphics[width=0.45\textwidth]{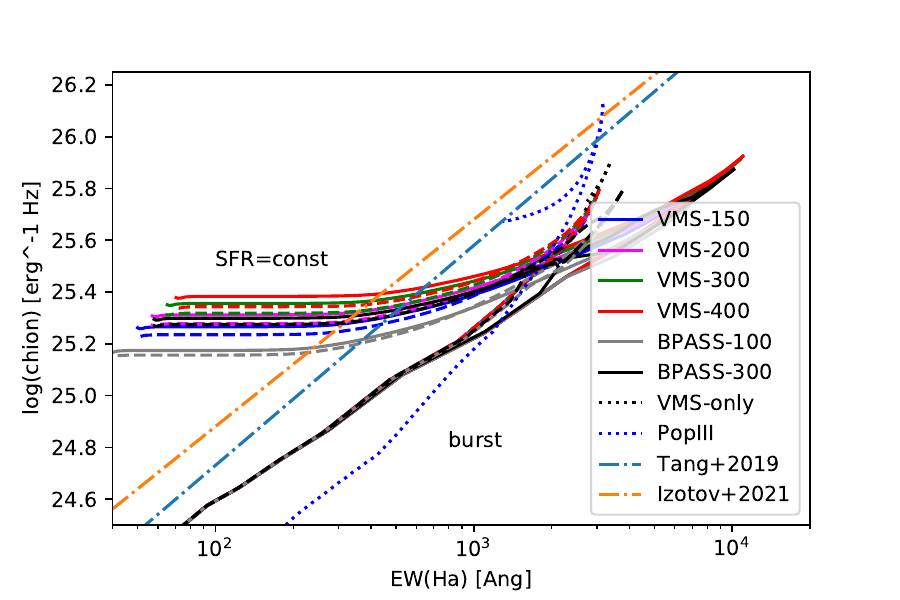}
\caption{Ionizing photon efficiency \chion\ as a function of age (left panel) and \ha\ equivalent width (right panel) of different models and star formation histories.
Solid (dashed) lines show models without (with) nebular continuum emission.
In the right panel we also show for comparison predictions for metal-free (Pop III) populations extending up to 500 \msun\ (blue dotted line).
Dash-dotted lines show the ``canonical'' value of \chion\ used by \cite{Robertson2013New-Constraints} (left panel), and the mean relation between \chion\ and EW(\ha) derived by 
\protect\cite{Tang2019MMT-MMIRS-spect}  and \cite{Izotov2021Low-redshift-co} from observations. }
\label{fig_chion}
\end{figure*}

\subsection{Properties of the SEDs in the Lyman continuum} 
\label{s_hardness}
 Both the strength of the Lyman break and the detailed shape of the SEDs in the Lyman continuum are altered by the presence of VMS as illustrated in Fig.~\ref{fig_seds}.
 To quantify the strength of the Lyman break we show the ratio of the ionizing to non-ionizing continuum flux at 1500 \AA, $L_{900}/L_{1500}$ measured in $F_\nu$ units, in Fig.~\ref{fig_lyb}. In models including VMS the LyC flux increases, in other words the break decreases, by factors up to $\sim 1.4$ at the youngest ages and $\sim 1.5$ for constant SFR. Accounting for nebular continuum emission, these effects lead to a global increase in $L_{900}/L_{1500}$ by no more than 30\% when VMS are included.
Earlier predictions of the break (L900/l1500), derived from \bpass\ or other models, are shown, for example, in \cite{schaerer2003,Siana2007New-Constraints,Steidel2018The-Keck-Lyman-} and they overall agree with the standard \bpass\ models excluding VMS shown here. 
 
To illustrate the shape of the ionizing radiation fields predicted by the different models we show their ionizing photon fluxes above a certain energy, $Q(>E)$, as a function of energy in Fig.~\ref{fig_qe}. All models are normalized at the Lyman limit (13.6 eV) for comparison, and the ionization potential of several important ions are indicated in the figure.
Clearly the SED shapes of all models are essentially undistinguishable between 13.6 and $\sim 35$ eV and they start to diverge at higher energies. The surprising result, a priori, is that the ionizing spectrum of stellar populations including VMS is softer and not harder as could naively be expected when more massive hot stars are added. This behavior is due to physical processes in the atmospheres of these stars, where the presence of strong (dense) stellar winds causes a different ionization structure resulting in softer spectra. In particular He$^{2+}$ recombines in the outer parts of these atmospheres, similarly to the case of WR stars discussed already by \cite{Schmutz1992Theoretical-Con}, leading, for example, to strong and broad \Heiiuv\ emission, which is in some way the main observational characteristics of VMS. Therefore, most ionizing photons above 54 eV are absorbed in the atmosphere of these stars, which explains the strong decrease of the ionizing flux at these energies.

Differences in the shape/hardness of the ionizing radiation field have implications for the relative strengths of nebular emission lines.
Since the SEDs are essentially identical for energies below 35 eV the relative intensities of main strong emission lines in the optical range (e.g., \Oii, \Oiii, \Nii, \Sii), which have ionization potentials between 13.6 and 35 eV, are expected to be basically invariant with respect to these differences. Only for lines of higher ionization stages, such as \civ, N~{\sc iv}, O~{\sc iv}, and \heii\ one expects changes, namely lower intensities compared, for example, to hydrogen lines or to other lines from lower ionization stages. We have indeed confirmed this qualitative behavior with calculations of photoionization models using {\em CLOUDY} \citep{Ferland2017The-2017-Releas}, which show that the strong optical line ratios are basically unchanged by the inclusion of the VMS models. Higher ionization lines are predicted to be weak in all models, with intensities most likely below the detection threshold of most observations. Therefore, the predicted changes due to VMS are probably difficult to detect, and we defer
a more thorough analysis of the impact of VMS on nebular emission line properties to  later.

\subsection{Ionizing photon efficiency}
An important quantity measuring the ionizing photon emission of star-forming galaxies is the so-called ionizing photon efficiency, $\chion = \qh / L_{\nu,1500}$, which measures the ionizing photon flux per unit monochromatic UV luminosity at 1500 \AA, $L_{\nu,1500}$, in units of erg s$^{-1}$ Hz$^{-1}$. The predictions for the different models considered here are shown in Fig.~\ref{fig_chion}, both including and excluding nebular continuum emission. 
At young ages, \chion\ is very high and the presence of VMS boosts the ionizing photon efficiency by up to a factor 1.4-1.5 (0.15-0.17 dex, for $\mup=400$ \msun). For constant SFR at equilibrium (ages $\ga 300$ Myr) \chion\ increases by up to $\sim 0.2$ dex when VMS are included, reaching $\log(\chion) \approx 25.4$ erg$^{-1}$ Hz, which is a factor 60\% higher than the ``canonical'' value used by \cite{Robertson2013New-Constraints}, derived from synthesis models from \cite{BC03}.

From previous studies it is well know that the ionizing photon efficiency of an integrated stellar population not only depends on age, star formation history and the IMF, but also on metallicity \citep[see e.g.,][]{schaerer2003,raiter2010,Eldridge2022New-Insights-in}.
In general, \chion\ is maximum at young ages ($\la 2$ Myr), increases for IMFs favoring massive stars (e.g., flatter slopes $\alpha_2$ and or larger maximum stellar masses \mup), and increases toward low metallicities.
The highest ionizing photon efficiencies reported from model predictions and ``normal'' IMFs (Salpeter slope and $\mup \sim 100-120$ \msun) are $\log(\chion)=25.55$ erg$^{-1}$ Hz for constant SFR and 26.05 at young ages for metal-free stellar populations \citep{raiter2010}. 
For a flatter IMF ($\alpha_2=-2.0$) \cite{Eldridge2022New-Insights-in} report a maximum of $\log(\chion)=25.55$ erg$^{-1}$ Hz for constant SFR. For more extreme IMF including only massive stars, the ionizing efficiency can reach up to $\log(\chion)=26.1$ erg$^{-1}$ Hz for metal-free stars, as shown by \cite{raiter2010}, and illustrated on the right panel of Fig.~\ref{fig_chion}.

It should be noted that there is a natural upper limit to \chion\ from stars/stellar populations, since the emission in the LyC and the non-ionizing UV are related (e.g., described by a Planck function or a more realistic SED) and since increasing LyC emission (\qh) will also lead to increased nebular continuum emission ($L_{1500}$ in particular). As Fig.~\ref{fig_chion} shows, the models including VMS up to 300-400 \msun\ show quite similar ionizing efficiencies at young age, with values $\log(\chion) \approx 25.8$ erg$^{-1}$ Hz when nebular continuum emission is accounted for, close to the maximum value of $\log(\chion) \approx 25.9$ erg$^{-1}$ Hz predicted by the most extreme models (VMS-only).
In short, the comparison of our  models with these extreme values clearly show that stellar populations at moderately low metallicity can already reach very high ionizing photon efficiencies during a short burst phase. In fact, if stellar populations including only VMS -- that is devoid of stars below $\sim 100$ \msun\ -- exist, their efficiency will be $\log(\chion) \approx 25.8-25.9$ erg$^{-1}$ Hz at LMC metallicity, within a factor of $\sim 2$ of the maximum possible efficiency of PopIII stars. 
The upper limit on \chion\ and other properties are further discussed in Sect.~\ref{s_discuss}.

In the right panel of Fig.~\ref{fig_chion} we show \chion\  as a function of the predicted \ha\ equivalent width, together with two average relations derived from observations of star-forming galaxies at low-redshift and at $z \sim 1.3-2.4$ \citep{Tang2019MMT-MMIRS-spect,Izotov2021Low-redshift-co}.  Although these relations are plotted here beyond the range over which they were established, one clearly sees that our models are shifted toward too high EW(\ha), even with the inclusion of nebular  continuum emission. Most likely this shift is due to the presence of some previous/older stellar populations in the observed galaxies, which reduces the observed EW(\ha). A contribution of the order of $\ga 50$\% to the optical continuum is already sufficient to reconcile the two. Alternatively, or in addition, the observationally derived \chion\ values could also be overestimated if the UV attenuation of these galaxies is systematically underestimated, although there is currently no indication for this \citep[see e.g.][]{Tang2019MMT-MMIRS-spect}.

\begin{figure*}[htb]
\centering
 \includegraphics[width=0.45\textwidth]{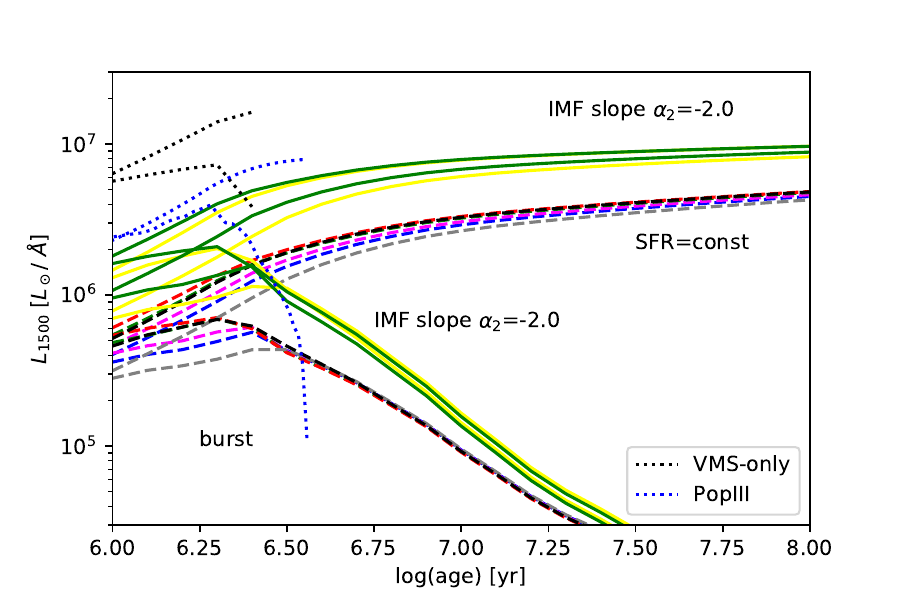}
 \includegraphics[width=0.45\textwidth]{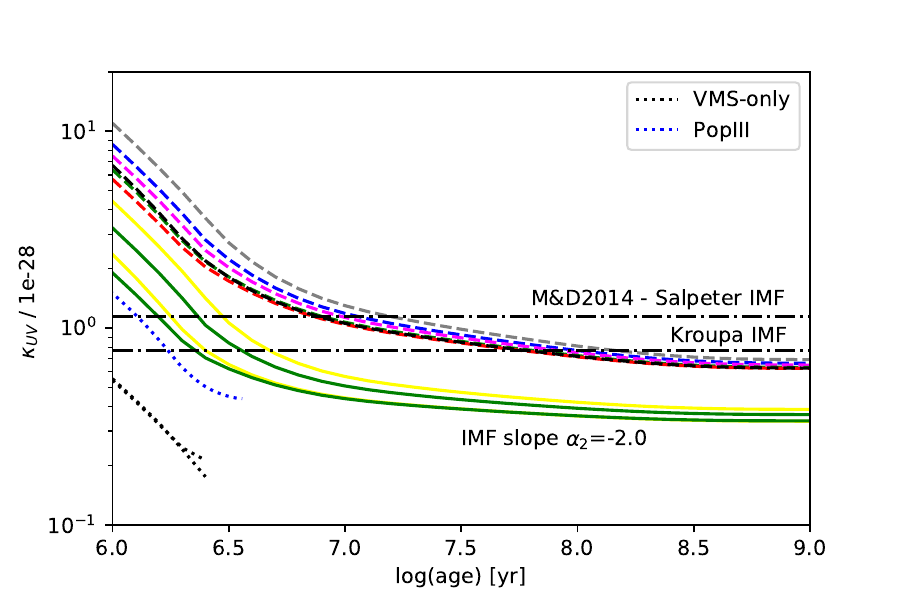}
 \caption{Temporal evolution of the monochromatic UV luminosity $L_{1500}$  (left) and the UV star formation rate conversion factor \kappauv\ (right panel, for SFR=const) predicted for models with a Salpeter IMF slope ($\alpha_2=-2.35$) and a flatter IMF ($\alpha_2=-2.$).  $L_{1500}$  is normalized to $10^6$ \msun\ for bursts and to 1 \msunyr\ for SFR=const. The models with a Salpeter slope are shown using the same colors as in previous figures. Models with the flat IMF, including also nebular emission), are shown by yellow and green lines. VMS-only models and PopIII models with a top-heavy IMF are shown by dotted lines. All models shown here include nebular  continuum emission. 
Dash-dotted horizontal lines in the right panel show the value of the  canonical SFR(UV) conversion factor from \cite{Madau2014Cosmic-Star-For} for a Salpeter and a Kroupa IMF.}
\label{fig_kappa}
\end{figure*}

\subsection{The UV luminosity of populations with VMS}
One of the most direct and obvious implications of the presence of VMS is to ``boost'' the UV luminosity of stellar populations. For example, in the LMC cluster R136 the seven most massive stars -- with masses between 110--325 \msun\ -- contribute 32 \% of the UV continuum flux at 1500 \AA, as shown by  \cite{Crowther2016The-R136-star-c}.  
In the left panel of Fig.~\ref{fig_kappa} we show the temporal evolution of the monochromatic luminosity $L_{1500}$ for all the models from Table \ref{ta_models}, both for instantaneous bursts and constant star formation. For comparison we also show predictions for a ``top-heavy'', metal-free population, using the PopIII models from \cite{raiter2010} with an IMF containing only stars between 50--500 \msun.   

As Fig.~\ref{fig_kappa} shows, for bursts the UV luminosity peaks around $\sim 2$ Myr before it decreases, typically by one order of magnitude over $\sim 10$ Myr. For constant SFR the UV luminosity progressively builds up with time, reaching an equilibrium after $\ga 50-100$ Myr. 
For populations with ``extreme'' IMFs, containing only VMS, this equilibrium is reached already at $\sim 3$ Myr, and further evolution is not shown here.
At ages $\la 3$ Myr, an increase in the maximum stellar mass, \mup\, leads to a higher UV luminosity, with boosts up to a factor $\sim 2$ when VMS up to 400 \msun\ are included. At later times the boost factor of the UV luminosity is smaller, only of the order of 1.14 at $\sim 100$ Myr for constant SFR. For bursts there is no difference after $\sim 3$ Myr, the lifetime of VMS.

The slope of the IMF has a strong impact on the resulting UV luminosity of integrated stellar populations, as shown in Fig.~\ref{fig_kappa}. Indeed, a small flattening of the IMF with a change of slope from Salpeter ($\alpha_2=-2.35$) to  $\alpha_2=-2.0$ already increases the UV luminosity significantly.
At young ages, for example, models with VMS up to 400 \msun\ and $\alpha_2=-2.0$ are brighter by a factor 5--6 than ``classical'' models
($\mup=100$ \msun,  $\alpha_2=-2.35$).
For models with constant SFR the boost in UV luminosity is a factor $\sim 2.2$ at ages $\ga 10^8$ yr.
Models including only VMS show, quite naturally, even higher UV luminosities per unit stellar mass or SFR, as also shown Fig.~\ref{fig_kappa}. 
For constant SFR, $L_{1500}/$SFR then surpasses the models with a fully populated IMF with $\alpha_2=-2.0$ even further, by an additional factor $\sim 2$. 
Overall, between the  UV luminosity predicted by standard BPASS models with $\mup=100$ \msun\ and the maximum for a ``VMS-only'' population, there is a factor $\sim$ 17 increase in non-ionizing UV output.
It is also interesting to note that metal-free populations PopIII do not have a higher UV luminosity than a VMS-only population at $Z=0.006$ at these wavelengths, since their average temperature is hotter and hence a larger fraction of the flux is emitted at higher energies (in the Lyman continuum).
The maximum values predicted by the models shown here are summarized in Table \ref{ta_predict}.

The resulting conversion factor between SFR and UV luminosity, \kappauv, defined as
\begin{equation}
	{\rm SFR(UV)} = \kappauv \times L_{\nu,1500}, 
\end{equation}
where $L_{\nu,1500}$ is in conventional units of \ergs\ Hz$^{-1}$, is shown in the right panel of Fig.~\ref{fig_kappa}.
The canonical values of $\kappauv = 1.15 \times 10^{-28}$ \kappaunits\ taken from \cite{Madau2014Cosmic-Star-For} for a Salpeter IMF from 0.1--100 \msun\ and the one adjusted to the Kroupa IMF used here ($\kappauv = 0.77 \times 10^{-28}$ in the same units, after scaling with a factor 0.67), are also shown in the figure for reference.
As this figure shows, all of our models with a standard IMF slope ($\alpha_2=-2.35$) yield SFR conversion factors which are very close to the canonical values from the literature. To first order the effect of an extension of the IMF to VMS has therefore a small effect on the SFR(UV) calibration factor  \kappauv.
However, as discussed before, the IMF slope has a strong impact on \kappauv, which is reduced by a factor $\sim$ 2.4 to $\kappauv = 0.32 \times 10^{-28}$ \kappaunits\ for a slope of $\alpha_2=-2.0$. Furthermore, the timescale over which this equilibrium value is reached is also strongly reduced to ages of $\ga 20-30$ Myr and the canonical value of \kappauv\ is already reached after $\sim 4-10$ Myr, as shown in Fig.~\ref{fig_kappa}.
Finally, models including only VMS (from 100--400 \msun) have the highest UV luminosity per unit SFR, yielding thus the lowest values found here with $\kappauv \approx 0.2 \times 10^{-28}$ \kappaunits. 

As for \chion, it is of interest to examine if there is a lower limit to the SFR(UV) conversion factor, or equivalently if there is an upper limit to the UV luminosity per unit SFR. The answer is indeed yes,  and this limit is essentially given by the maximum UV luminosity $L_{1500}$/\mstar\ per unit mass of massive stars. Now to first order $L/\mstar$ is constant for stars above $\mstar \ga 100-150$ \msun\ \citep[see e.g.,][]{Martins2020Spectral-proper},  and their effective temperature at the zero-age main sequence also, which implies that $L_{1500}/\mstar \approx$ constant with stellar mass. Therefore, increasing \mup\ and/or flattening the IMF slope will not further increase the UV luminosity of a stellar population of fixed stellar mass or  SFR beyond a certain limit. The VMS-only models shown here illustrate this limit for \kappauv\ and \chion\ also, for moderately low metallicities.

We note that the value of \kappauv\ depends on metallicity but in a non-monotonic fashion, as has already been discussed by \cite{raiter2010}.
Indeed, while the UV luminosity per unit stellar mass increases first from solar to low metallicities due to the increase of the average effective temperature of stars on the main sequence, $L_{1500}$ decreases below $\la 1/50$ solar since the bulk of the flux of very hot stars is then emitted at wavelengths $\ll 1500$ \AA. For this reason the minimum value of $ \kappauv = 0.18 \times 10^{-28}$ \kappaunits\  obtained here for $Z=0.006$ for a very ``top-heavy'' IMF is even lower than for top-heavy IMFs at zero metallicity\footnote{Note that the UV SFR conversion factors \kappauv\ show by \cite{Harikane2023A-Comprehensive} for PopIII models (taken from \cite{Zackrisson2011The-Spectral-Ev} who use the same models as the present study, from \cite{raiter2010}) have erroneously been corrected by a factor 1.49, and are therefore too low by this factor.}.

\begin{figure}[htb]
\centering
 \includegraphics[width=0.45\textwidth]{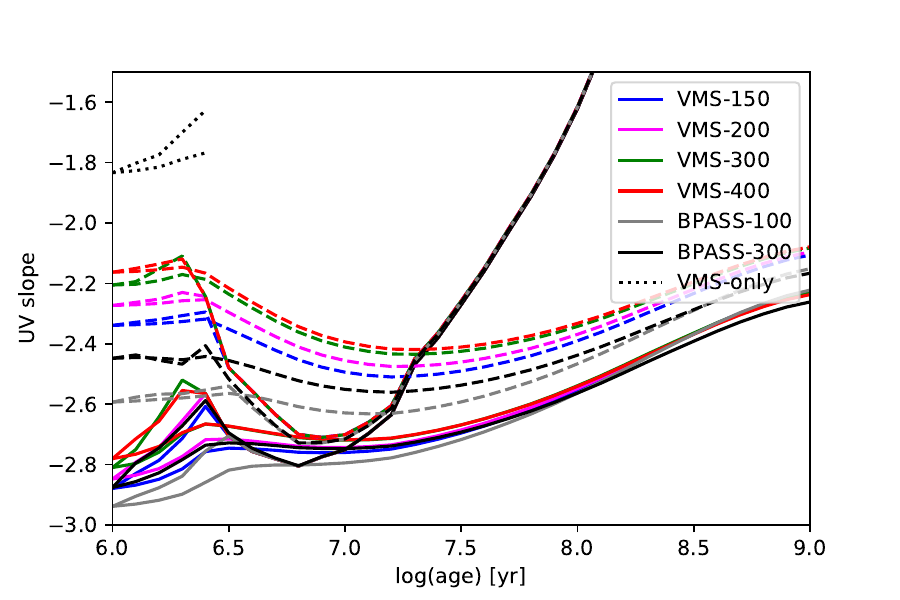}
\caption{UV slope $\beta_{1550}$ as a function of age for the main models presented in this work.  Solid lines show the slopes derived for pure stellar models, neglecting nebular continuum emission. Dashed and dotted lines show $\beta_{1550}$ including nebular continuum emission.}
\label{fig_beta}
\end{figure}

\subsection{The UV slope}

The predicted UV slopes of our models, $\beta_{1550}$ measured here between  1300--1800 \AA\ following \cite{raiter2010}, are plotted in Fig.~\ref{fig_beta}, for models with and without nebular continuum emission, and for the two star formation histories. The effect of VMS can be appreciated at the youngest ages ($\la 3$ Myr) and in models without nebular emission, where we see that the UV slope becomes somewhat redder (i.e., $\beta$ increases) when VMS are included and with increasing \mup.This is due to the average effective temperature of the VMS, which is somewhat lower than that of stars with $\sim 100$ \msun, as shown in \cite{Martins2022Spectroscopic-e}. In other words with increasing stellar mass the zero-age main sequence does not become hotter in the VMS regime.
This behavior of the VMS leads to a non-monotonous evolution of  $\beta_{1550}$ at young ages ($\la 5$ Myr).
The resulting changes in $\beta$ are of the order of $\Delta \beta \sim 0.2-0.3$, which is relatively small, and comparable to the uncertainties of many measurements of the UV slope at high-$z$. 

As is well known, the nebular continuum changes significantly the UV slope, making it redder by $\Delta \beta \sim 0.3-0.6$ at young ages \citep[e.g.,][]{schaerer2003,raiter2010}.
For models with constant SFR, the effect of VMS changes the UV slope by up to $\sim 0.1$ after $\sim 10$ Myr, and by smaller amounts over longer timescales.  Again, the effect of nebular continuum emission, in other words including nebular emission or not, is more important than an extension of the IMF into the VMS regime.

Interestingly, for the reasons already mentioned above, the UV spectral slope of populations including or even dominated by VMS are not extremely blue, as one could naively have expected. For example, the stellar spectrum of populations hosting only VMS  have $-2.6 \la \beta_{1550} \sim -2.0$ (without nebular continuum). On the contrary, when nebular emission is added, VMS-only spectra become significantly redder, with $\beta_{1550} \sim -1.8$. In practice, this means that UV slope should not be a good diagnostic for the presence/absence of VMS and other features should be examined for this purpose.
 

\begin{figure}[htb]
\centering
 \includegraphics[width=0.45\textwidth]{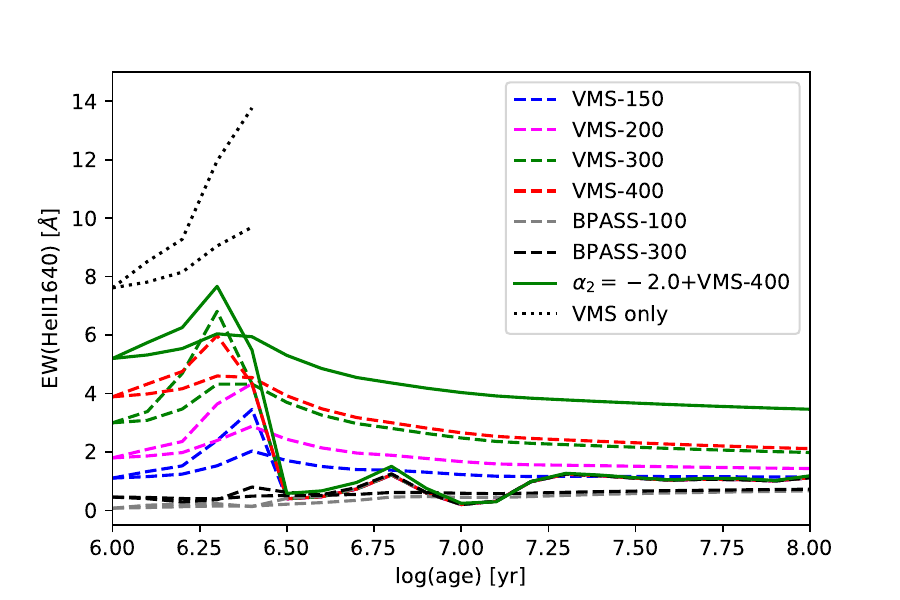}
\caption{Predicted equivalent width of the stellar \heii\ emission as a function of age for the different models.}
\label{fig_heii}
\end{figure}

\subsection{Direct signatures of VMS}

Several observational features can potentially be used to detect the presence of VMS from integrated spectra 
\citep[see e.g., discussions in][]{Hadfield2006How-extreme-are,Wofford2014A-Rare-Encounte,Smith2016The-Very-Massiv,Smith2023HST-FUV-Spectro,Martins2023Inferring-the-p,Crowther2024Mapping-the-cor}. From our current knowledge of individual VMS studied in the Galaxy and the LMC,  the clearest and strongest feature of VMS in the UV is their strong and broad \Heiiuv\ emission line, which can significantly outshine the emission from Wolf-Rayet stars in this line \citep{Martins2023Inferring-the-p,Upadhyaya2024Evidence-for-ve}.
On the other hand, \cite{Crowther2024Mapping-the-cor} showed empirically that the contribution from classical Wolf-Rayet stars in mixed-aged populations (NGC 2070 including R136) can be significant ($\sim 56$\%). 
There is, however, a concensus that existing stellar population models essentially all fail to reproduce the main features of observed UV-spectra of low-metallicity objects with strong \Heiiuv\ emission without inclusion of VMS \citep[see][]{Senchyna2021Ultraviolet-spe,Smith2023HST-FUV-Spectro,Martins2023Inferring-the-p,Crowther2024Mapping-the-cor}.

The predicted strength of the \heii\ line from our models is shown in Fig.~\ref{fig_heii}, where the equivalent widths have been derived from the BPASS+VMS high resolution SEDs using the same spectral windows as in the study of  \cite{Upadhyaya2024Evidence-for-ve}. Our results agree with theirs, but we present additional models accounting for nebular continuum emission and exploring other IMFs.
A maximum EW$($\heii$) \sim 7$ \AA\ is predicted for models with a classical Salpeter IMF and $\mup= 400$ \msun\ and the EW decreases with decreasing maximum stellar mass \mup. Steepening the IMF slope increases the maximum EW at young ages to $\sim 8$ \AA\ and leads to EW$\sim 4-5$ \AA\ for constant SFR after $\ga 10-100$ Myr.
Finally, the most ``extreme'' models considered here, where only VMS are present, predict \heii\ equivalent widths between $\sim 8-14$ \AA\ depending on age and star formation history (bursts or constant SFR), as also shown in Fig.~\ref{fig_heii}.

\section{Discussion}
\label{s_discuss}

\begin{table*}
\caption{Maximum values predicted for bursts and values at equilibrium for models with constant SFR} 
\label{ta_predict}
\begin{tabular}{lccccc|ccccc}
Model &  \multicolumn{5}{c|}{Burst ($10^6$ \msun)} & \multicolumn{5}{c}{constant SFR ($=1$ \msunyr)}  \\
ID & $\log(Q)$ & EW(\hb) & $\log(\chion)$ & $L_{1500}$ & $M_{UV}$ & $\log(Q)$ &$\log(\chion)$ & $L_{1500}$ & $M_{UV}$  & \kappauv \\
    & [s$^{-1}$] & [\AA] & [erg$^{-1}$ Hz] & [$\lsun\AA^{-1}$] & & [s$^{-1}$] & [erg$^{-1}$ Hz] & [$\lsun\AA^{-1}$]  & & \\
\hline 
\multicolumn{5}{c}{Predictions including nebular emission:}\\
BPASS-100 & 52.65 & 532 & 25.69 & 4.36e+05 & -16.16 & 53.32 & 25.15 & 5.13e+06 & -18.84 & 6.75e-01 \\
BPASS-300 & 52.93 & 654 & 25.79 & 6.89e+05 & -16.66 & 53.48 & 25.27 & 5.65e+06 & -18.94 & 6.13e-01 \\
VMS-150 & 52.81 & 530 & 25.74 & 5.90e+05 & -16.49 & 53.43 & 25.23 & 5.45e+06 & -18.90 & 6.35e-01 \\
VMS-200 & 52.90 & 557 & 25.77 & 6.39e+05 & -16.57 & 53.48 & 25.27 & 5.58e+06 & -18.93 & 6.21e-01 \\
VMS-300 & 52.99 & 579 & 25.79 & 7.34e+05 & -16.73 & 53.53 & 25.31 & 5.74e+06 & -18.96 & 6.04e-01 \\
VMS-400 & 53.04 & 588 & 25.80 & 7.50e+05 & -16.75 & 53.56 & 25.34 & 5.81e+06 & -18.97 & 5.96e-01 \\
Flatter-IMF-150 & 53.29 & 592 & 25.79 & 1.64e+06 & -17.60 & 53.86 & 25.41 & 9.87e+06 & -19.55 & 3.51e-01 \\
Flatter-IMF-400 & 53.56 & 636 & 25.84 & 2.22e+06 & -17.93 & 54.02 & 25.53 & 1.07e+07 & -19.64 & 3.23e-01 \\
VMS-only-150 & 54.17 & 712 & 25.91 & 9.89e+06 & -19.55 & 54.61 & 25.82 & 2.13e+07 & -20.38 & 1.63e-01 \\
VMS-only-400 & 54.15 & 689 & 25.89 & 7.79e+06 & -19.29 & 54.54 & 25.83 & 1.76e+07 & -20.17 & 1.97e-01 \\
PopIII (IMF C) & 54.05 & 753 & 26.18 & 3.94e+06 & -18.55\ & 54.439 & 26.08  &7.90e+06 & -19.30 & 0.438\\
\multicolumn{5}{c}{Predictions without nebular continuum emission:}\\
BPASS-100 & 52.65 & 832 & 25.76 & 4.03e+05 & -16.07 & 53.32 & 25.17 & 4.93e+06 & -18.79 & 7.03e-01 \\
BPASS-300 & 52.93 & 1173 & 25.87 & 6.25e+05 & -16.55 & 53.48 & 25.29 & 5.35e+06 & -18.88 & 6.47e-01 \\
VMS-150 & 52.81 & 1067 & 25.85 & 5.08e+05 & -16.32 & 53.43 & 25.26 & 5.09e+06 & -18.83 & 6.81e-01 \\
VMS-200 & 52.90 & 1183 & 25.88 & 5.46e+05 & -16.40 & 53.48 & 25.30 & 5.16e+06 & -18.84 & 6.71e-01 \\
VMS-300 & 52.99 & 1286 & 25.91 & 6.01e+05 & -16.51 & 53.53 & 25.35 & 5.27e+06 & -18.86 & 6.57e-01 \\
VMS-400 & 53.04 & 1332 & 25.92 & 6.08e+05 & -16.52 & 53.56 & 25.38 & 5.31e+06 & -18.87 & 6.52e-01 \\
Flatter-IMF-400 & 53.56 & 1609 & 25.98 & 1.77e+06 & -17.68 & 54.02 & 25.60 & 9.26e+06 & -19.48 & 3.74e-01 \\
VMS-only-150 & 54.17 & 2209 & 26.08 & 7.92e+06 & -19.31 & 54.61 & 25.95 & 1.57e+07 & -20.05 & 2.21e-01 \\
VMS-only-400 & 54.15 & 2002 & 26.06 & 6.05e+06 & -19.01 & 54.54 & 25.97 & 1.29e+07 & -19.83 & 2.69e-01 \\
\end{tabular}
\tablefoot{Maximum values predicted for bursts (columns 2-6) and values at equilibrium for models with constant SFR (cols.~7-11).
Predictions in the top-half of the table include nebular continuum emission, those below correspond to pure stellar emission.
We also list the predictions for PopIII models with a top-heavy IMF (IMF C) from \protect\cite{raiter2010} for comparison.
\kappauv\ is given in units of $10^{-28}$ \kappaunits. Burst models are normalized to a total mass of $10^6$ \msun, constant star formation to 
SFR$=1$ \msunyr. ``Flatter-IMF-X" refers to models with a slope $\alpha_2=-2.0$ and an upper mass cut-off $\mup=$X.}
\end{table*}

\subsection{Comparison with other studies}
A detailed comparison between different models that allow for extensions of the IMF beyond $\ga 100$ \msun\ is beyond the scope of this paper. 
We only briefly mention several studies which discuss UV and ionizing properties of such populations and quickly comment on some aspects of interest.

The work of  \cite{raiter2010} presents extensive predictions for 8 different IMFs including in some cases VMS up to 500 \msun, covering a wide range of metallicities from PopIII ($Z=0$) to solar metallicity. However, for metallicities $Z \ge 1/50$ \zsun\ the upper mass limit is 120 \msun. 
\cite{Stanway2019Initial-mass-fu} present \bpass\ models for a wide range of metallicities, $\mup=100$ and 300 \msun, and two IMF slopes (identical as those used here). Their paper mostly focuses on the hardness of the ionizing spectra and discusses how the hardness depends on IMF variations. 
Our study differs from this work in that it uses tailored atmosphere models for VMS, which leads to softer spectra for VMS compared to those adopted in their models,
for the reasons discussed in Sect.~\ref{s_hardness}. Our study is not optimized to examine in detail the hardness of the ionizing spectra, since we neglect post main-sequence phases which can contribute to emission at higher energies, as included in the work of \cite{Stanway2019Initial-mass-fu}. Overall, their results include those presented as our BPASSS-100 and BPASS-300 models.

\bpass\ predictions for different IMFs are also show in \cite{Eldridge2022New-Insights-in}. The \chion\ values reported there (for the metallicity $Z=0.006$ used here) are in good agreement with our values, when the same assumptions are made.
Independent predictions of the ionizing photon production $Q$ and \kappauv\ for different IMFs and metallicities have been presented in \cite{Lapi2024Constraining-th}, who use these data to constrain the IMF during the epoch of reionization from comparisons to UV-luminosity functions and other observations. Overall their results appear to differ from ours. While their canonical value for $Q$ (for the Chabrier IMF) agrees with our predictions, our top-heavy (VMS-only) IMFs predict significantly higher values of $Q$. Their maximum $Q$ values are comparable to ours for $\alpha_2=-2.35$ and $\mup=400$ \msun, which would indicate that despite the vast parameter space they explore, their model does not predict large enough variations in the  ionizing photon production. Also,  their standard value of \kappauv\ (which is defined as $1/\kappauv$) seems incompatible with ours. Finally, their predictions for very top-heavy IMF (with characteristic masses $m_c \ga 100$ \msun\ seem incoherent, as they indicate lower UV and ionizing photons emission than for normal IMFs.  In short, we find it difficult to understand their predictions.

\subsection{Maximum UV and ionizing photon production of stellar populations and their implications}

The maximum UV and ionizing photon emission from stellar populations is of interest to understand, e.g., the UV-bright high-$z$ galaxies observed with the JWST, and in other contexts.
The predictions from our models, exploring a range of IMFs, and including also for comparison the most extreme metal-free stellar population 
 \citep[IMF C from][a top-heavy IMF  with stars between 50--500 \msun]{raiter2010}, are listed in Table \ref{ta_predict}. 

Overall the table shows the following:
First, for a young simple stellar population (burst) with a Salpeter IMF with different upper mass limits (up to 400 \msun),  the ionizing photon production ($Q$) increases by up to a factor 2 and the rest-UV luminosity (here at 1500 \AA) by up to a factor 1.6, whereas the maximum ionizing photon efficiency remains essentially constant, for reasons discussed earlier.
Significantly stronger boosts are obtained with a flatter IMF, where a small change of the IMF slope (for example, changing $\alpha_2=-2.35$ to  $\alpha_2=-2.0$) can lead to enhancements in the maximum ionizing photon emission by a factor 4--8 and a boost of the UV luminosity by a factor $\sim 4.8$ (1.7 magnitude).
Considering stellar populations exclusively made of VMS ($\mstar \ga 150$ \msun), one obtains maximal values which are basically independent of the exact IMF (slope and upper mass limit), and which are $\sim 30$ times higher for the ionizing photon flux and  $\sim 21$ times higher (3.3 magnitude brighter) for the UV luminosity, both per unit stellar mass. These are ``absolute'' maximal values (or a minimum UV mass-to-light-ratio) limited by fairly basic principles of stellar physics, which can therefore not significantly be surpassed by other stellar populations, irrespective of their metallicity and IMF. This also includes metal-free stellar populations (Pop III), as comparison with the model from \cite{raiter2010} with the most extreme IMF shows.
Only the maximum of \chion\ and the hardness of the ionizing spectrum, measured for example by  ratio of the ionizing photons emitted above 54 eV (the ionization potential of He$^+$) compared to those above 13.6 eV, are higher for PopIII (for the reasons discussed earlier), exceeding the maximum values ``VMS-only'' populations.

Second, for constant star formation rates and assuming that one reaches an equilibrium after $\ga 10-100$ Myr depending on the quantity considered,
the ionizing photon production ($Q$) and \chion\ increase by up to a factor 1.7, whereas the rest-UV luminosity is boosted by only $\la 10$ for a fully populated Salpeter IMF ($\alpha_2=-2.35$) , extending down to the low-mass regime. Larger boosts, of a factor $\sim 2$, are obtained with a flatter IMF ($\alpha_2=-2.0$). The maximum ionizing photon emission per unit SFR, obtained for populations consisting exclusively of VMS, is a factor $\sim 20$ higher than with a classical IMF. For this case, the UV luminosity is boosted by a factor $\sim 4$, implying a fairly fundamental value of min(\kappauv)$\approx 1.8 \times 10^{-29}$ \kappaunits\ for the mimimum  of the SFR(UV) conversion factor.

Finally, we also include for completeness the predictions neglecting  nebular continuum emission in Table \ref{ta_predict}. In this case the UV emission at 1500 \AA\ is decreased, the corresponding quantities are lower, and both \chion\ and EW(\hb) higher. In general and for the interpretation of integrated observations of star-forming galaxies in particular, nebular emission should be present, and hence models including it should preferentially be used. Values intermediate between the predictions with and without nebular emission are expected, for example if a fraction of the ionizing photons escape or are destroyed by dust inside the \hii\ regions.

The ``boost'' in UV luminosity and ionizing production per unit stellar or star formation rate, has obviously potentially important implications for our understanding of star-forming regions and galaxies, and it could contribute to understanding or solving some of the puzzles mentioned in the introduction.  As our models show, only small differences in the IMF slope have significant effects on the UV luminosity and ionizing photon production. And the presence of VMS further boosts these effects. Could the IMF be steeper and/or extend beyond the classical mass limit of $\sim 100$ \msun\ ? There is various evidence supporting both possibilities, at least in some environments/conditions.
%
Individual stars with masses in the range of $\sim 100-300$ \msun\ have been found in the compact young star cluster R136 in the 30 Dor region
\citep{Crowther2010The-R136-star-c,Bestenlehner2020The-R136-star-c,Kalari2022Resolving-the-C}. The presence of VMS is also suspected or established in several low-$z$ star-forming regions  \citep[e.g.,][]{Leitherer2018Physical-Proper,Smith2016The-Very-Massiv,Senchyna2021Ultraviolet-spe,Wofford2023Extreme-broad-H,Smith2023HST-FUV-Spectro}, in a lensed star cluster at $z=2.37$ \cite{Mestric2023Clues-on-the-pr}, and in UV-bright galaxies at $z \sim 2-4$ \cite{Upadhyaya2024Evidence-for-ve}.

The IMF slope of 30 Doradus, which includes R136, the prototypical nearby region with VMS, has extensively been studied. Surrounding the central region with R136, where crowding limited their analysis,
\cite{Schneider2018An-excess-of-ma} found indications for a flatter-than-Salpeter IMF, with a slope $\alpha=-1.90^{+0.37}_{-0.26}$ (in our units, where the Salpeter slope is $\alpha=-2.35$), and an observed number of VMS in R136 compatible with this IMF.
And many other studies indicate that the IMF could be flatter (at least for massive stars) in dense gas environments and/or at low metallicity 
\citep[e.g.,][]{Dabringhausen2009A-top-heavy-ste,Marks2012Evidence-for-to,Dib2023Variation-of-th}
Similarly at integrated galaxy scales, several studies have proposed a flatter or ``top-heavy'' IMF in galaxies with high star formation rates
\citep[e.g.,][]{Weidner2011Top-heavy-integ}  and in sub-mm galaxies \citep[cf.][]{Baugh2005Can-the-faint-s,Zhang2018Stellar-populat}, although this result is disputed
\cite[see e.g.,][]{Safarzadeh2017Is-a-top-heavy-}. Other works have combined a variety of different constraints including supernova rates, the extragalactic background and others, concluding, however, that it is difficult to distinguish a Salpeter IMF from steeper ones \citep[cf.][]{Ziegler2022Non-universal-s}.
Finally, further examples are works who examine possible shifts of the characteristic mass of the IMF with redshift, using detailed measurements of galaxy SEDs from large surveys. \cite{Sneppen2022Implications-of} and \cite{Steinhardt2022Implications-of}, for example, suggest that the IMF shifts to somewhat higher characteristic masses with increasing redshift,  becoming thus more top-heavy (or equivalently bottom-light), albeit with the same Salpeter slope for massive stars. Another recent approach to constrain the IMF from UV luminosity functions and observations probing the cosmic reionization history considers a more flexible description of the IMF, finding that IMF slopes flatter than Salpeter are excluded and favoring also a higher characteristic mass \citep[see][]{Lapi2024Constraining-th}. Numerous other studies have also speculated about a non-universal IMF at high redshift \citep{Trinca2024Exploring-the-n,Cueto2024ASTRAEUS.-IX.-I,Woodrum2023JADES:-Using-NI}, although no consensus exists on this question and clear observational tests are difficult to find.

\subsection{Observational tests for extreme IMFs}

\subsubsection{Possible observational tests}
Our spectral modeling of populations including VMS and considering non-standard IMFs can offer some hints on how to search for ``extreme'', very top-heavy IMFs. The simplest signatures offering direct tests would be: {\em 1)}  strong stellar \Heiiuv\ emission, {\em 2)} very high ionizing photon efficiencies \chion, {\em 3)} high $L(\ha)/L({\rm UV})$ ratios,  {\em 4)} and/or H recombination lines with very high equivalent widths, as we shall now discuss and compare with recent JWST results.

The most top-heavy IMFs considered here, those including only VMS stars, can be tested quite easily and already excluded for many objects. Indeed, in such a case all galaxies should show very strong \Heiiuv\ emission (with EW $\ga 8$ \AA), if the model predictions for the metallicity considered here are applicable, which we discuss below. Furthermore  the ionizing photon efficiency and other quantities such as the \ha\ or \hb\ EWs would be limited to very high values ($\log(\chion) \ga 25.7$ erg$^{-1}$ Hz), since such populations would contain only short-lived and VMS. The fact that \chion\ typically covers a relatively wide range including median values significantly lower than the above value excludes already such extreme IMFs for the majority of star-forming galaxies from of high to low-$z$  \citep[see e.g.,][for measurements of \chion\ from $z \sim 0-7$ ]{Bouwens2015Using-the-Infer,Izotov2021Low-redshift-co,Prieto-Lyon2023The-production-}. 

Many other observations, including the color distribution of galaxies, the presence of Balmer break and others, testify to the presence of lower mass, longer or long-lived stars in most galaxies, which therefore exclude very bottom-light IMFs as a frequent or universal phenomenon. 
This is also the case for certain galaxies at high redshift, where Balmer breaks have been established with JWST both from photometry and spectroscopy \citep[see e.g.,][]{Carnall2023A-massive-quies,Carnall2024The-JWST-EXCELS,Looser2024A-recently-quen}.
As is well known, the Balmer break, for example, probes A-type stars with temperatures $\teff \sim 10$ kK, which corresponds to main-sequence turn-off masses $\mstar \sim 2-3$ \msun\ \citep[with lifetimes $\tau_H \sim 300-700$ Myr for metallicities $(0.1-1)\times$ solar;][]{Ekstrom2012Grids-of-stella}. 
If the presence of  stars down to 2--3 \msun\ is established it implies that $\sim 40$ \% of the total mass of the population is probed, assuming a Kroupa/Chabrier IMF, which means that the UV luminosity per unit mass cannot be boosted by more than a factor $\sim 2.5$.

In short, very bottom-light IMFs or extremely top-heavy IMF can be probed observationally, and IMFs with purely VMS are already excluded for bulk of high-$z$ galaxies.
Now, if we assume that the IMF is populated with stars over a wide mass range (e.g., \mlow $\ll 2-3$ \msun, and $\mup \sim 100$ \msun\ or higher) constraining the IMF, its exact slope and upper mass limit, is more challenging, as we shall now examine.  

\subsubsection{Comparisons with JWST observations}
More specifically, we shall now discuss if there is some direct evidence for unusual IMFs and/or the presence of VMS in the high-$z$ galaxies observed with JWST.

From stacked JWST spectra of galaxies from $z \sim 5-11$ \cite{Roberts-Borsani2024Between-the-Ext} find that \Ciiiuv\ is the strongest line with EW$\sim 5-14$ \AA\ (increasing with $z$).  This excludes already that these galaxies have on average a VMS-only IMF, since \Heiiuv\ is weaker (or non-detected) than \Ciiiuv, whereas VMS-only models predict EW(\heii) $\sim 8$ \AA. On the other hand, the data may just be compatible with a flatter IMF and \mup $\sim 400$ \msun (our Flat-IMF+VMS-400), on average.

Then, individual galaxies should be better tests of VMS and extreme IMF, since the signatures of VMS (which have short lifeftimes) are best identified in objects dominated by young populations.
 \cite{Castellano2024JWST-NIRSpec-Sp} recently discovered a very bright galaxy at $z=12.34$ showing numerous strong emission lines, including \Heiiuv. The possible detection of this line (albeit of low significance) with EW=$4.9 \pm 3.1$ \AA\ is in the range of our models with VMS and a flat IMF. However, the line could also trace nebular emission (fully or partly) and there is thus no clear evidence for VMS in this object. 
\cite{Wang2024A-Strong-He-II-} reported the possible detection of \Heiiuv\ in a blue galaxy at $z=8.16$, suggesting the possible presence of Pop III stars  co-existing with metal-enriched stellar populations. According to these authors \heii\ is strong, with an equivalent width EW(\Heiiuv)$=21\pm4$ \AA, and dominated by nebular emission, although the width of the line shown in their Fig.~3 appears relatively broad and comparable to observation by \cite{Upadhyaya2024Evidence-for-ve} which indicate the presence of VMS. However, the strength of \heii\ clearly surpasses those of previous observations in other objects and also model predictions for normal and extreme stellar populations. Further studies of this peculiar object would be warranted.
Beyond these detections, most spectroscopic studies of high-$z$ galaxies with JWST do not detect \Heiiuv\ and provide only relatively loose upper limits. For example, 
 \cite{Carniani2024Spectroscopic-c} have observed two $z  \sim 14$ galaxies, detecting \Ciii\ emission but no \heii, with upper limits EW(\Heiiuv)$<7$ (16) \AA\ (3 $\sigma$) for the two objects. Similarly,  for $z \sim 10$ galaxies \cite{Curtis-Lake2023Spectroscopic-c} report upper limits of EW(\Heiiuv)$<13.5-15.4$ \AA\ for three galaxies (one has $<6$ \AA\, all 2 $\sigma$ limits). Clearly these data do not exclude the presence of VMS, flatter IMFs, and even the VMS-only scenario. Better, high S/N, medium resolution spectra are needed to provide tighter constraints.

Two interesting objects with clear nebular continuum emission and strong lines have recently been reported by \cite{Cameron2024Nebular-dominat}. The authors suggest that these galaxies are ``completely dominated'' by nebular emission and would require a very top-heavy IMF.  We find it unlikely that the galaxy GS-NDG-9422 at $z=5.943$ which they discuss in detail, is dominated by nebular continuum emission, since the observed equivalent widths of the H recombination lines are significantly lower than the predicted values assuming a pure nebular continuum. Indeed, in this case one has $EW = \epsilon_{ij} / \gamma_{\rm tot}$, where $ \epsilon_{ij}$ is the  line emissivity and  $ \gamma_{\rm tot}$ the continuous emission coefficient at the wavelength of the line. For \hb, for example, one has $EW(\hb) \sim 1000$ \AA\ for typical electron temperatures and assuming Case B, as shown in Fig.~\ref{fig_ew_nebular}. This value exceeds the observed equivalent width of GS-NDG-9422 ($EW(\hb) \approx 400$ \AA) by a factor $\sim 2.5$, showing that the nebular continuum contributes approximately $\sim 40$ \% of the total in the rest-optical.  Also a boost of the two-photon continuum emission is accompanied by enhanced \ha/\hb\ ratios \citep{raiter2010}, opposite to the low Balmer decrement observed  in this galaxy \cite{Cameron2024Nebular-dominat}.
Furthermore, the observed ionizing photon efficiency, $\log(\chion) \approx 25.7$ erg$^{-1}$ Hz as obtained from the \hb\ flux and the observed UV continuum flux, falls short of the theoretical efficiency  $\log(\chion) \approx 26.3-26.4$ erg$^{-1}$ Hz for pure nebular continuum  emission (see Appendix). 
Other sources of emission must therefore contribute, as has already been pointed out  by \cite{Tacchella2024Resolving-the-n} and \cite{Li2024No-top-heavy-st}, who show that the spectrum of GS-NDG-9422  can be explained with a combination of nebular, stellar, and AGN emission, without the need for unusual stellar populations \cite[see also][]{Heintz2024The-JWST-PRIMAL}. 

Several studies have measured the ionizing photon production of high-$z$ galaxies with JWST from photometry, by determining $Q$ from the \ha\ luminosity, which can be obtained from ``flux'' excesses detected in specific filters, or in a more indirect fashion. For example, \cite{Prieto-Lyon2023The-production-} and \cite{Simmonds2024Low-mass-bursty} find median values of  $\log(\chion) =25.31 \pm 0.43$   and $\log(\chion) > 25.5$ erg$^{-1}$ Hz at $z>6$, respectively, with some objects reaching $\log(\chion) \approx 26.0$ erg$^{-1}$ Hz, close to the maximum predicted by our models.
The median values are clearly higher than those predicted for standard IMFs at equilibrium; that is for constant SFR over sufficiently long timescales. However, this does not imply that IMFs favoring massive, or VMS are needed, if most of objects are dominated by populations younger than 10--50 Myr.
According to our models $\log(\chion) > 25.7$ erg$^{-1}$ Hz requires either VMS (masses $>$ 100 \msun) and or an IMF flatter than Salpeter. A fraction of the galaxies observed by   \cite{Prieto-Lyon2023The-production-} and \cite{Simmonds2024Low-mass-bursty} show indeed such high ionizing efficiencies, indicating possible IMF differences. Alternate explanations are an overestimate of \chion\ if dust attenuation was underestimated, or lower metallicities on average. However, since the maximum of \chion\ only increases by $\sim 0.1$ dex when metallicity decreases from solar to 1/100 solar \citep[cf.][]{raiter2010} this is insufficient to explain galaxies with $\log(\chion) \ga 25.8$ erg$^{-1}$ Hz with a standard IMF. A more in-depth analysis of the high-\chion\ objects could be of interest to understand if this is related to a peculiar IMF, or to additional but not accounted for ionizing sources (e.g.~AGN). However, we note that the objects with the highest \chion\ values are generally not UV-bright, which suggests that the latter are not strongly boosted in the UV and ionizing emission beyond a standard IMF.


Since by definition $\chion = Q_H/L_{\nu,1500} = (L(\ha)/c_{\ha})  (\kappauv/L_{\nu,1500})$, where $L(\ha)$ is the \ha\ luminosity and $c_{\ha}=\epsilon_{32}/\alpha_B$ (cf.~Sect.~\ref{s_appendix}),  the ionizing photon efficiency is also proportional to $L(\ha)/L(UV)$ and the two quantities depend in the same way on the IMF slope and upper mass limits.
Indeed, measurements of  $L(\ha)/L(UV)$ have long been used to measure and constrain the upper end of the IMF using observations of star clusters and integrated galaxy spectra \citep[see e.g.,][and references therein]{Meurer2009Evidence-for-a-,Lee2009Comparison-of-H,Bastian2010A-Universal-Ste,Ashworth2017Exploring-the-I}. Alternatively, $L(\ha)/L(UV)$ or analogous measurements of SFR(\ha)/SFR(UV) are often used to examine the ``burstiness'' or stochasticity of star formation \citep[e.g.,][]{Dominguez2015Consequences-of,Emami2019A-Closer-Look-a,Atek2022The-star-format}, since this quantity also depends on the star formation history. Since $L(\ha)/L(UV)$ and \chion\ are equivalent, searches for IMFs favoring massive or VMS would therefore also look for objects exceeding the maximum $L(\ha)/L(UV)$ predicted by standard models, as was already discussed above.

In principle, if the range between the maximum value of $L(\ha)/L(UV)$ (or \chion)  at young ages and the value at equilibrium can be measured this can also be used to constrain the IMF, as can be seen from Table \ref{ta_predict}.  Whereas for IMFs with a Salpeter slope these quantities decrease by $0.48-0.54$ dex from maximum to equilibrium, this range is reduced for flatter IMFs (reaching e.g., 0.32 dex for Flat-IMF-400), down to $\sim 0.1$ dex for VMS-only populations. The latter is already excluded for the majority of objects, which show $\log(\chion) < 25.7$ erg$^{-1}$ Hz from the JWST measurements. 
 If stacking of multiple objects provides an average corresponding to constant SFR, measurements from stacked spectra could be compared to the maximum value (determined e.g., from individual objects) for this test. If this is feasible in practice remains, however, to be seen.
Three recent studies have constructed stacked spectra for galaxies  between $z \sim 3$ and 10 from JWST NIRSpec observations \citep{Roberts-Borsani2024Between-the-Ext,Langeroodi2024NIRSpec-View-of,Kumari2024JADES:-Physical}.  
For all galaxies \cite{Roberts-Borsani2024Between-the-Ext} measure $\log(\chion)=25.30 \pm 0.01$ erg$^{-1}$ Hz, in good agreement with the median $\log(\chion)=25.33 \pm 0.47$ erg$^{-1}$ obtained by \cite{Prieto-Lyon2023The-production-}, and with our  equilibrium predictions for constant SFR with a standard IMF extending to $\sim 200-300$ \msun. From the data of \cite{Prieto-Lyon2023The-production-} the observed difference between their maximum $\log(\chion) \approx 26.0$  erg$^{-1}$ Hz and this value is $\sim 0.7$ dex, larger than the value predicted by the models for all IMFs considered here. Possibly this is due to uncertainties in the observational determination of \chion, which are of the order of 0.1-0.2 dex or larger, and which can be overestimated if the dust attenuation is underestimated.  Again, more in-depth studies and possibly additional/more data would be needed to constrain the IMF.

Finally, the last method to test for IMFs favoring massive stars or VMS is to search for objects with high equivalent widths of hydrogen lines (e.g., \ha, \hb),
which, if exceeding some maximum value for standard IMFs (say EW$(\hb) \sim 550$ \AA\ or even $\ga 800$ \AA\ if nebular continuum emission can somehow be avoided), would require non-standard IMFs. In principle this test is straightforward, although the amount of time during which EWs exceed this limit are short, less than $\la  2-3$ Myr at best, and therefore only very few objects are expected in such a phase. Other possibilities to encounter such high EWs include observations where stellar emission is spatially offset from the nebular emission region, in which case maximum EWs can reach very high values, as illustrated by the pure nebular case in Fig.~\ref{fig_ew_nebular}.

To summarize, several  features accessible with rest-UV-to-optical observations can in principle be used to constrain the IMF or more specifically test if extreme IMFs showing deviations from the Salpeter slope and/or extending to significantly higher masses than commonly adopted.  However, since these features also depend on other parameters, most importantly star formation histories, age and metallicity, some of which are difficult to constrain,  these tests are not straightforward in practice. Whereas top-heavy IMF are therefore difficult to constrain, extremely bottom-light IMFs -- for example populations consisting only of (very) massive stars which would be very UV-luminous per stellar mass -- are probably fairly simple to exclude.

\subsection{Caveats}
\label{s_caveats}
Our predictions build on the \bpass\ synthesis models of \cite{Eldridge2017Binary-Populati} including binary stars for a metallicity $Z=0.006$, which are combined with the state-of-the-art stellar evolution and atmosphere models for VMS from  \citep{Martins2022Spectroscopic-e} which have successfully been tested against observations of individual VMS in the LMC (R136). Our combined models have also been compared against other observations of low-redshifts star-forming regions and galaxies with metallicities comparable to the LMC \citep{Martins2023Inferring-the-p}. Finally, they have also successfully been applied to the model the Sunburst arc spectrum, a strongly lensed star cluster at $z=2.37$, and they reproduce the observed rest-UV spectra of very UV-bright galaxies at $z \sim 2-4$ that show signatures pointing toward the presence of VMS  \citep{Mestric2023Clues-on-the-pr,Upadhyaya2024Evidence-for-ve}, although Wolf-Rayet features have been found in the Sunburst arc by \cite{Rivera-Thorsen2024The-Sunburst-Ar}.

Despite these successes and confrontations with observations our models have also some limitations. First, the VMS models used here follow only their main-sequence evolution, neglecting thus later evolutionary stages, primarily for computational reasons.
In general post main-sequence evolution is typically $\sim 10$\% of the total stellar lifetime, which implies that the impact of these phases should overall be small or limited. 
The line intensity of \Heiiuv\ depends mostly on luminosity, mass loss, helium mass fraction, and ionization (thus mostly \teff). The most advanced models computed by \cite{Martins2022Spectroscopic-e} already have more than 80\% of mass in the form of helium. Their mass loss rates are high, of the order $10^{-4}$~\msunyr, which is the maximum observed for normal Wolf-Rayet stars in the LMC \citep{hainich15,aadland22}. The mass loss recipe of \cite{graef21} do not predict significantly larger values for VMS. Finally, the luminosity of post main-sequence VMS in the LMC should be within 0.2 dex of that on the MS \citep[e.g.,][]{Martinet2023Very-Massive-St}. Consequently we do not expect a quantitative boost of the \Heiiuv\ intensity in the latest phases of evolution, at least not strong enough to compensate for the reduced duration of these phases compared to the main-sequence. 
For comparison, the VMS models of \cite{Yusof2013Evolution-and-f} have a phase of very strong mass loss, which leads to the formation of WR stars after core H-burning accompanied by a rapid decrease in luminosity and an evolution to higher temperatures. We therefore primarily expect that such objects would increase the ionizing flux at high energies but not strongly alter the stellar \Heiiuv\ emission which will be dominated by the more luminous VMS during H-burning. However, combined stellar evolution and atmosphere models including also these phases will be needed to properly quantify these statements.
In any case, we note that from normal Wolf-Rayet stars we typically expect \heii\ emission reaching EW(\Heiiuv) $\approx 2-3$ \AA\ during a short phase at $\sim 4-5$ Myr, according to the models of \cite{SV98} at $Z=0.008$, and that none of the existing stellar synthesis models with normal IMFs (i.e., without VMS)  predict strong enough \Heiiuv\ lines to explain the observations \citep[see also][]{Wofford2014A-Rare-Encounte,Smith2016The-Very-Massiv,Smith2023HST-FUV-Spectro,Martins2023Inferring-the-p,Crowther2024Mapping-the-cor}.

Second, the present work discusses only models at a single metallicity $Z=0.006 \approx 0.44$ solar, since detailed, well-calibrated VMS models treating consistently both stellar evolution and atmosphere, are so far only available for this metallicity. This means that it is somewhat uncertain how our results can be transposed to other metallicities, in particular to lower metallicities of prime interest for high-$z$ studies.
Indeed, since several quantities, in particular the strength of the \Heiiuv\ emission line, are very sensitive to the stellar wind properties of VMS, the main unknown is probably how mass loss depends on metallicity for VMS, a topic currently under investigation \citep{sander23,sabhahit23}.
Interestingly, from observations of Mrk 71, a star-burst region with $\sim 16$\% solar metallicity showing strong stellar \Heiiuv\ emission, and comparisons with R136 and other higher-metallicity regions, \cite{Smith2023HST-FUV-Spectro} find that the \Heiiuv\ emission does not significantly change with metallicity and therefore suggest that the mass loss properties of VMS should be similar at low metallicity.
First results from Martins et al. ~(in preparation) indeed suggest that this main VMS signature remains present and strong, at least down to 0.1 solar metallicity.
Then, the metallicity-dependence of the properties of massive stars (i.e., UV luminosity, ionizing photon production etc.) is accounted for in standard synthesis models (BPASS and others), and, as discussed above, metallicity effects appear overall smaller than changes of the IMF, such as an extension toward VMS and/or changes of the IMF slope. 
We therefore expect that the relative effects of VMS on stellar populations are, to very first order at least, similar and independent of metallicity between $\sim 0.1-0.5$ solar. Future work will be needed to examine this in detail and more quantitatively.

\section{Conclusion}
\label{s_conclude}

Combining the stellar evolution and atmosphere models of VMS from \cite{Martins2022Spectroscopic-e} for a moderately-low metallicity 
($Z=0.00536=1/2.5$ \zsun) with \bpass\ evolutionary synthesis models from \cite{Eldridge2017Binary-Populati} we have computed extensive sets of spectral energy distributions for integrated stellar populations hosting VMS with masses between 150 and 400 \msun.  This work expands on the previous models and analysis of UV spectra of nearby star-forming regions, the Sunburst arc at $z=2.37$, and UV-bright galaxies at $z \sim 2-4$ presented earlier by  \cite{Martins2022Spectroscopic-e}, \cite{Mestric2023Clues-on-the-pr}, \cite{Martins2023Inferring-the-p}, and \cite{Upadhyaya2024Evidence-for-ve}.

We considered two limiting cases of star formation histories, instantaneous bursts and constant star formation and explored a wide range of IMFs, including different upper mass limits, and an IMF slope between Salpeter ($\alpha_2=-2.35$) and flatter ($\alpha_2=-2$). We also considered stellar populations with extreme IMFs, for example some consisting only of VMS stars with initial masses between 100--400 \msun. Accounting also for nebular continuum emission and H recombination lines, we predict integrated SEDs and derived quantities, such as the UV luminosity (\luv), ionizing photon production ($Q$) and efficiency (\chion), the strength of the Lyman break, H line luminosities and equivalent widths, the UV slope, and the strength of the \Heiiuv\ line (which is dominated by VMS); the main quantities are presented in  Table \ref{ta_predict}. The main predictions from our paper can be summarized as follows:
\begin{itemize}
\item VMS contribute significantly to the UV luminosity and Lyman continuum emission of young stellar populations: for example, for IMFs with a Salpeter slope the inclusions of stars up 400 \msun\  boost the luminosity at 1500 \AA, \luv, by a factor of 1.63 at young ages and by $\sim 11$ \% at equilibrium, for constant SFR over long timescales. Similarly, the ionizing photon production, $Q$, is boosted by factors of 2.45 (for young bursts) and 1.74 (constant SFR).
Larger boosts are obtained with flatter IMFs, and only small changes in the IMF slope (adopting $\alpha_2=-2$ instead of the Salpeter slope) lead to increases by a factor of 4.8 (8.1) at young ages and 2 (5) for SFR=const for the non-ionizing and ionizing UV, \luv\ ($Q$) .

\item The LyC emission of VMS leads to an increase in the ionizing photon efficiency, \chion, by 0.14-0.2 dex for young bursts or constant SFR and a Salpeter slope. For $\alpha_2=-2$, the boost reaches up to 0.18-0.4 dex when the IMF is extended up to 400 \msun. Increased LyC emission also leads to stronger emission lines, with H recombination line luminosities boosted by the same factors as $Q$, and equivalent widths reaching $EW(\hb) \sim 700$ \AA\ at the youngest ages.

\item The SEDs including VMS show reduced Lyman breaks. The shape of the ionizing spectra of populations with or without VMS is similar, between the LyC limit and up to $\sim 35$ eV. At higher energies the spectra are softer with VMS, due to the increased opacity in the winds of VMS, caused by the boosted mass loss rate they experience compared to normal massive stars. The differences in ionizing spectra do not affect the classical, optical emission line (so-called BPT) diagrams.

\item Populations with VMS show strong stellar  \heii\ emission, with EW(\Heiiuv) up to 4--8 \AA\ at young ages or $\sim 2.5-4$ \AA\ for constant SFR, considering IMF slopes between  $\alpha_2=-2.35$ and $-2$. (Fig.~\ref{fig_heii}).

\item Extreme IMFs populated only by stars in the VMS regime lead to ``extreme'' values for most of observational properties considered here.
Per unit stellar mass (for young bursts) the UV luminosity ($Q$) can be boosted by a factor $\sim 20$ (30) approximately. For constant SFR at equilibrium the maximum boost (compared to a classical Kroupa/Chabrier IMF extending to 100 \msun) is $\sim 4$ (16) for \luv\ ($Q$) by unit SFR.
The ionizing photon efficiency of such populations is $\log(\chion) \sim 25.85-25.95$ erg$^{-1}$ Hz, and even $\sim 0.1$ dex higher if nebular continuum emission is absent. 
These values are essentially ``fundamental'' maximum limits, which are independent of the exact IMF slope and maximum stellar mass. This is the case since stellar models in the VMS domain follow $L \propto M$ and show similar zero-age main sequence  properties. They are therefore to first order scale-free, and populations consisting purely of VMS are therefore indistinguishable, that is insensitive to the detailed shape and upper mass limit of the IMF, and only dependent on the total amount of stars formed as VMS.
Compared to very top-heavy IMFs considered sometimes for metal-free populations, our moderately metal-poor VMS-only models emit $\sim 2$ times more UV photons per stellar mass or SFR in the non-ionizing UV (1500 \AA\ for example), since the average temperature is less extreme than those of Pop III, whose emission peaks well in the Lyman continuum. This comparison illustrates the relatively weak metallicity dependence of the predictions of such extreme populations \citep[see also][]{raiter2010}.
 
\end{itemize}

Regarding the need to explain UV-bright galaxies in the early Universe (see Sect.~\ref{s_intro}), we found that the addition of VMS in the range of 100--400 \msun\ provides only relatively modest boosts of the UV luminosity, if the IMF slope remains Salpeter.  However, small changes of the IMF slope can significantly increase the UV and ionizing flux output of galaxies.

Finally, we have presented several methods to observationally search for unusual IMF which would favor (very) massive stars. We showed how measurements of the \Heiiuv\ line, \chion, $L(\ha)/L({\rm UV})$, and H recombination lines can constrain the high mass tail of the IMF, and also how these quantities and others, such as the Balmer break, can be used to exclude ``bottom-light'' IMFs.
Using published JWST observations we have examined if high redshift ($z \ga 5-6$) galaxies show some evidence of the presence of VMS and/or signs of non-standard IMFs. Extreme IMFs containing only VMS can be excluded on average, but the current data does not exclude IMFs with Salpeter slope or somewhat steeper (e.g.. $\alpha_2=-2$) including VMS for the majority of galaxies. Rest-UV spectra with higher spectral resolution and better S/N are needed to establish the existence of VMS in high-$z$ galaxies.
Furthermore, consistent stellar evolution and atmosphere models for VMS at lower metallicities than those currently available will be needed to examine if and how ``universal'' their spectral features are, and to quantify more broadly their impact.


The results from our models are made available in electronic format on Zenodo\footnote{\url{https://doi.org/10.5281/zenodo.14176049}}.
It is our hope that the present study will help researchers to better understand the stellar content and IMF of the most distant galaxies and their impact on the ISM, cosmic reionization, and other related topics.

\bibliographystyle{aa}
\bibliography{merge_misc_highz_literature}
 \begin{appendix}
\section{Nebular continuum and H recombination lines}
\label{s_appendix}

\begin{figure}[htb]
\centering 
 \includegraphics[width=0.45\textwidth]{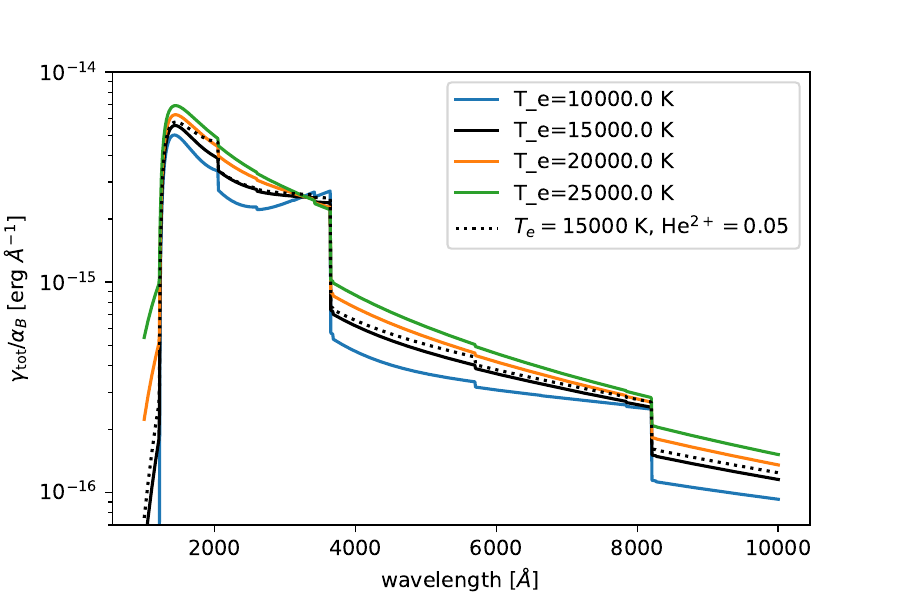}
\caption{Nebular continuum emission $\gamma_{\rm total}/\alpha_B$ of H plus He (with standard abundances) as a function of wavelength and for different electron temperatures. The default value adopted here is shown in black. The dotted line shows $\gamma_{\rm total}/\alpha_B$ for a highly ionized mixture with $n({\rm He}^+)/n({\rm H}^+)=n({\rm He}^{2+})/n({\rm H}^+)=0.05$. }
\label{fig_nebcont}
\end{figure}

\begin{figure}[htb]
\centering 
 \includegraphics[width=0.45\textwidth]{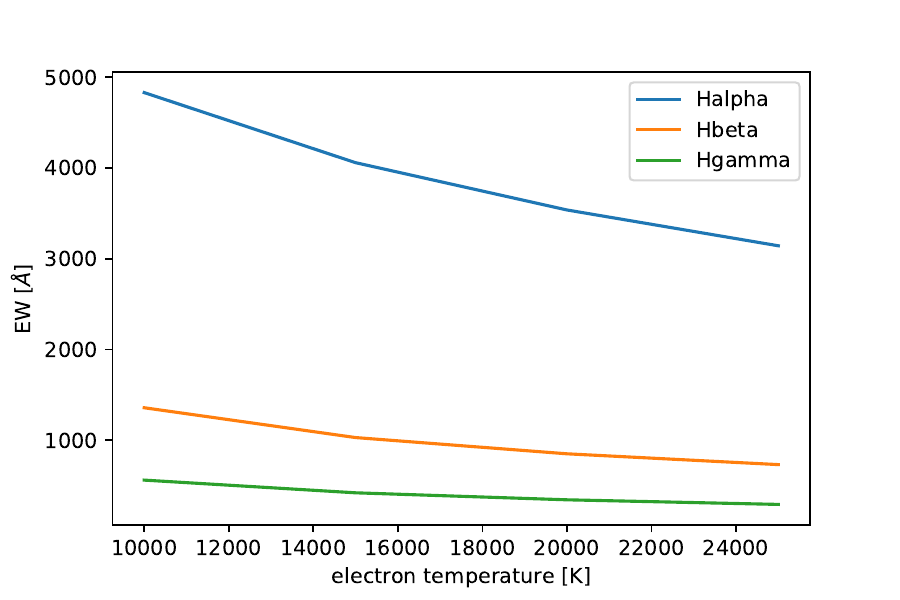}
\caption{Predicted equivalent width of  the equivalent widths of \ha, \hb, and H$_\gamma$ for a nebular continuum as function of the electron temperature}
\label{fig_ew_nebular}
\end{figure}

We briefly illustrate the properties of the nebular continuum adopted in the present work and discuss several simple properties related to nebular continuum emission and H recombination lines. All predictions have been computed using {\tt pyneb} \citep{Morisset2020Atomic-Data-Ass}.

The monochromatic luminosity of the gas is given by
\begin{equation}
  \label{eq_lcont}
  L^{\rm neb}_\lambda = \frac{\gamma_{\rm total}}{\alpha_B} (1 - f_{\rm esc}) Q({\rm H}),
\end{equation}
where $\alpha_B$ is the case B recombination coefficient for hydrogen. The continuous emission coefficient $\gamma_{\rm total}$, including free-free and free-bound emission by H and He, as well as the two-photon continuum of hydrogen is given by
\begin{equation}
  \label{eq_gamma}
  \gamma_{\rm total} = \gamma_{\rm HI} + \gamma_{\rm 2q}
        +  \gamma_{\rm HeI} \frac{n({\rm He}^+)}{n({\rm H}^+)}
        +  \gamma_{\rm HeII} \frac{n({\rm He}^{2+})}{n({\rm H}^+)}.
\end{equation}
The emission coefficients $\gamma_i$ (in units of erg cm$^3$ s$^{-1}$ \AA$^{-1}$) are taken from {\tt pyneb}. We assume a standard He abundance of 0.1 (in number) and that singly ionized He dominates (i.e.,  $n({\rm He}^{++})/n({\rm H}^+)=0.$). The resulting continuum emission is illustrated in Fig.~\ref{fig_nebcont} for different electron temperatures, $T_e$. The models discussed in this paper assume a typical value $T_e=15000$ K. Higher (lower) electron temperatures would result in somewhat lower (higher) values for the predicted \chion\ and equivalent widths of \ha\ and \hb. The effect of doubly ionized He, if present, should be negligible in most cases.

At metallicities $\la 0.03-0.1$ solar, well below those adopted here, the nebular continuum can in principle be enhanced compared to the predictions shown here. Such a boost is due to collisional effects, which overpopulate the $n=2$ level and enhance the photoionization rate, boosting thus the two-photon continuum emission by a factor which is proportional to the mean energy of the ionizing photons. At the same time the collisions also modify the Balmer decrement, increasing, for example, the \ha/\hb\ ratio \citep[see]{raiter2010}.  At the metallicity adopted here these effect are, however, negligible.

Significant increases of the  electron density beyond $n_e \ga 10^{3-4}$ \cmc\ decrease the two-photon emission, in other words nebular emission in the rest-UV, but has little effect at longer wavelengths. On the other hand, this collisional effect leads to a boost of \lya\ emission.
\citep[For a more detailed discussion of these effects, which may affect the nebular continuum and recombination line properties of individual objects at very low metallicity see e.g., ][]{Bottorff2006Two-Photon-Tran,raiter2010,Katz202421-Balmer-Jump-}.

By definition, the ionizing photon efficiency is
\begin{equation}
 \chion = \frac{\qh}{L_{1500}} = \frac{\qh}{L^\star_{1500}+ L^{\rm neb}_{1500}}
 \end{equation}
Hence, for a case where the stellar continuum can be neglected, one has  
\begin{equation}
\chion^{\rm neb} = \frac{\alpha_B}{\gamma_{\rm total} (1 - f_{\rm esc})}.
 \end{equation}
For electron temperatures between 10-20 kK one obtains $\log(\chion^{\rm neb})=26.3-26.4$ erg Hz$^{-1}$.
 
 For completeness, the fraction of energy contributed by nebular continuum emission at 1500 \AA, $f_{\rm neb}$,
can easily be inferred from 
\begin{equation}
  \label{eq_fneb}
  f_{\rm neb} = 1 - \frac{\chion}{ \chion^\star},
\end{equation}
where $\chion^\star = \qh / L^\star_{1500}$ is the ionizing photon efficiency neglecting nebular continuum emission,
whose values are reported at the bottom of Table \ref{ta_predict}.

Similarly, for H recombination lines the luminosity is
\begin{equation}
  \label{eq_hlines}
  L^{\rm ij}  = \frac{\epsilon_{ij}}{\alpha_B} (1 - f_{\rm esc}) Q({\rm H}),
\end{equation}
where $\epsilon_{ij}$ is the line emissivity. Hence the equivalent width, neglecting the stellar continuum becomes
\begin{equation}
  \label{eq_ew}
  EW^{\rm neb}({\rm ij})  = \frac{\epsilon_{ij}}{\gamma_{\rm total}}.  
\end{equation}
These values and their dependence on electron temperature are shown in Fig.~\ref{fig_ew_nebular}  for selected Balmer lines. For \hb, one has typically
EW$(\hb) \approx 1000$ \AA. These purely nebular EWs are, for example, convenient to infer the contribution of the stellar (and/or other) continuum sources from the measured EW of a H recombination line.

\end{appendix}

\end{document}